\documentclass[nofootinbib,aps,prd,10pt,superscriptaddress,showpacs,preprintnumbers,eqsecnum]{revtex4-1}

\usepackage{amsmath}
\usepackage{dsfont}
\usepackage{multirow}
\usepackage{graphicx}
\usepackage{mathrsfs}
\usepackage{amsfonts}
\usepackage{amssymb}
\usepackage[dvips]{color}

\begin{document}
\newcommand\be{\begin{equation}}
\newcommand\ee{\end{equation}}
\newcommand\bea{\begin{eqnarray}}
\newcommand\eea{\end{eqnarray}}
\newcommand\bseq{\begin{subequations}} %solo con amsmath
\newcommand\eseq{\end{subequations}}
\newcommand\bcas{\begin{cases}}
\newcommand\ecas{\end{cases}}
\newcommand{\p}{\partial}
\newcommand{\f}{\frac}

\title{Reflections on the hyperbolic plane}

\author{Orchidea Maria Lecian}
\email{lecian@icra.it}
\affiliation{Sapienza University of Rome, Physics Department, P. A. Moro, 5 - 00185 Rome, Italy}

%\today

\begin{abstract}
The most general solution to the Einstein equations in $4=3+1$ dimensions in the asymptotical limit close to the cosmological singularity under the BKL
(Belinskii- Khalatnikov- Lifshitz) hypothesis, for which space gradients are neglected and time derivatives only are considered, can be visualized by the behavior of a billiard ball in a triangular domain on the Upper Poincar\'e Half Plane (UPHP). The behavior of the billiard system (named 'big billiard') can be schematized by dividing the succesions of trajectories according to Poincar\'e return map on the sides of the billiard table, according to the paradigms implemented by the BKL investigation and by the CB-LKSKS (Chernoff- Barrow- Lifshitz- Khalatnikov- Sinai- Khanin- Shchur) one. Different maps are obtained, according to different symmetry-quotienting mechanisms used to analyze the dynamics according to the symmetries of the billiard domain and to the features of the geodesics on the UPHP. In the inhomogenous case, new structures have been uncovered, such that, in this framework, the billiard table (named 'small billiard') consists of $1/6$ of the previous one . The properties of the billiards described in these cases have already been studied in the reduced (according to the Poincar\'e return map) phase space.\\
The connections between the symmetry-quotienting mechanisms are further investigated on the UPHP. The relation between the complete billiard and the small billiard are also further explained according to the role of Weyl reflections.\\
The quantum properties of the system are sketched as well, and the physical interpretation of the wavefunction is further developped. In particular, a physical interpretation for the symmetry-quotienting maps is proposed, according to the Fourier decomposition of the energy levels of the wavefunction of the universe, as far as the meaning of epochs and eras is concerned from a quantum point of view.
\end{abstract}

\pacs{ 98.80.Jk Mathematical and relativistic aspects of cosmology- 05.45.-a Nonlinear dynamics and chaos;}

\maketitle

\section{Introduction\label{section1}}
The description of the chaotic behavior of a generic cosmological solution was approached in \cite{KB1969a}, \cite{Khalatnikov:1969eg}, \cite{BK1970}, \cite{BLK1971}, \cite{Lifshitz:1963ps} by locally reducing the corresponding Einstein equations to a system of ordinary differential equations (ODE's), and then by finding an approximated solution for the system of ODE's. In particular, this reduction is possible under the assumption that, close to the cosmological singularity, spacial gradients can be neglected with respect to time derivatives, such that points are spatially decoupled. The resulting behavior is one consisting of a succession of (suitably - joined) Kasner solutions. The statistical analysis of the model was achieved in \cite{KB1969a} and \cite{LLK}, where the one-dimensional BKL map (named after Belinskii, Khalatnikov and Lifshitz) was established. This statistical analysis was given a precise geometrical interpretation in \cite{Misner:1969hg}, \cite{chi1972}, \cite{Misner:1994ge} where the statistical properties of the BKL map were connected to the features of a billiard on 
a Lobachesky plane.\\
In \cite{Chernoff:1983zz}, \cite{sinai83} and \cite{sinai85}, the CB-LKSKS (named after Chernoff, Barrow, Lifshitz, Khalatnikov, Sinai, Khanin and Shchur) map is defined by upgrading to BKL map for a system of two variables. The BKL map is obtained from the CB-LKSKS map by marginalizing the 'extra' variable in the definition of probabilities associated to the map. The definition of all these maps is based on a suitable symmetry-quotienting mechanism. A link between two-variable maps and the discrete map for billiards in the Lobacesky plane was given in \cite{Kirillov:1996rd}.\\
The precise relations between the dynamics of the asymptotic Bianchi XI universe and its symmetry-quotiented versions (according to the identification of the role of Kasner exponents, and its symmetry-quotiented version according to the geometrical features of the billiard table) were established in \cite{Damour:2010sz}. This was possible by means of the identification of the geometrical expression of probabilities for billiards (given as the integral of an invariant reduced form over the pertinent domain in the reduced phase space) and the statistical probabilities for the statistical maps.\\
The link between the volume of the small billiard and that of the big billiard is given in \cite{Fleig:2011mu}.\\
Full information about cosmological billiards from different viewpoints can be found in \cite{Damour:2002et}, \cite{hps2009}, \cite{Montani:2007vu}.\\
\\
In this work, the features of the unquotiented billiard dynamics and those of its two symmetry-quotiented versions are defined for complex variables and analyzed according to their structure, implemented on the Upper Poincar\'e Half Plane (UPHP). In particular, the mechanisms of symmetry-quotienting the roles of the Kasner exponents is demonstrated to directly imply the proper boundary identifications of the group domains that generate the billiard maps on the UPHP. The maps for the reduced phase space given in \cite{Damour:2010sz} are immediately recovered as the limit on the absolute of the UPHP. The reduced-phase-space method \cite{Damour:2010sz} and the UPHP one are illustrated to be complementary for the definition of epochs and eras and their succession.\\
\\
The paper is organized as follows.\\
In Section \ref{section2}, the features of cosmological billiards are recalled. In particular, the main steps in the definition of cosmological billiards are outlined, and much attention is paid to the definitions of the unquotiented 'big billiard' and of its two possible symmetry-quotiented mechanisms, i.e. according to the permutation of the Kasner exponents and to the symmetry properties of the big-billiard table.\\
In Section \ref{section2a}, the basic tools in the definition of the groups are introduced, out of which the billiard maps are implemented in the UPHP for complex variables.\\
In Section \ref{section3}, the properties of these three kinds of billiards are analyzed on the UPHP. More in detail, a difference is found between quotienting out the role of the two oscillating Kasner exponents during a given BKL era, and identifying the 'newly-oscillating' Kasner exponent of the next BKL era with one of the previous BKL era. This difference implies a difference in the boundary identification for the fundamental domain where the two maps act. As a result, a paradigm to construct the unquotiented billiard map, the BKL map and the CB-LKSKS maps starting from the geometrical symmetry-quotienting of the unquotiented domain is established.\\
The kind of congruence subgroup which defines the big billiard with respect to the small billiard as far as cosmological billiards are concerned is discussed in Section \ref{comparison}.\\
The quantum regime is analyzed in Section \ref{section4}, where the properties of the wave function, resulting from the solution of the Hamiltonian constraint, are fixed according to the different physical features of the billiard table. More in detail, the symmetries of the billiard table and of the dynamics are taken into account, as far as the definition of the mathematical features of the wave function is concerned.\\
The role of (BKL) epochs and eras in the quantum version of the model is analyzed in Section \ref{section4a}, where the meaning of the 'quantum' BKL numbers is also investigated.\\
The direct comparison of these results with those found in \cite{BLK1971}, \cite{Chernoff:1983zz}, \cite{sinai83}, \cite{sinai85}, and \cite{Damour:2010sz} is accomplished in Section \ref{comparison}. The specification of \cite{Fleig:2011mu} is solved.\\
Brief concluding remarks in Section \ref{section5} end the paper.
%%%%%%%%%%%%%%%%%%%%%%%%%%%%%%%%%%%%%%%%%%%%%%%%%%%%%%%%%%%%%%%%%%%%%%%%%%%%%%%%%%   
\section{Basic Statements\label{section2}}
The metric of a general inhomogeneous space-time close to the cosmological singularity \cite{Damour:2002et} is suitably parametrized by the Iwasawa decomposition of the spacial metric, vanishing shift functions $N^i$ and an appropriate choice of the lapse function $N$. Close to the singularity, the limit of the Iwasawa decomposition of the spacial metric reads
\begin{equation}\label{iwa}
g_{ij}(x^0,x^i)=\sum_{a=1}^d e^{-2\beta^a(x^0,x^i)}N^a_{\ \ i}(x^0,x^i)N^a_{\ \ j}(x^0,x^i).
\end{equation}
The asymptotic limit close to the singularity
is characterized by the $\beta^a$ functions only, while the remaining degrees of freedom (given by the upper triangular matrices $\mathcal{N}^a_{\ \ i}(x^0,x^i)$) are frozen. The corresponding model is that of a massless particle moving in the ($\beta$) space with metric
\begin{equation}\label{betametric}
d\sigma^2=\sum_{a,b=1}^{d}G_{ab}d\beta^a d\beta^b=\sum_a\left(d\beta^a\right)^2-\left(\sum_ad\beta^a\right)^2
\end{equation}
in a closed domain delimited by infinite sharp walls. The dynamics is therefore that of a billiard, i.e. characterized by (elastic) 'bounces' against the walls and 'free-evolution' (geodesics) trajectories between the billiard walls (given by the asymptotic limit of the corresponding potential).\\ 
The free evolution between two bounces is given by the Kasner solution, the 'curvature' walls describe the asymptotic limit of the homogeneous Bianchi IX model, while the 'symmetry' walls have to be considered for inhomogeneous models (details will be given in the next paragraphs).\\
The Kasner metric is obtained for a diagonal metric with $g_{ij}=a(t)dx+b(t)dy+c(t)dz$, where $a$, $b$, $c$ are the Kasner scale factors, such that $a(t)\equiv e^{-\beta_1}$ (and analogous identifications for the other variables). It admits the solution $\beta^a=v^a\tau+v^a_0$, where the time variable $\tau$ is defined via $d\tau=-dt/\sqrt{g}$. The Kasner velocities $v^a$ define the Kasner parameters $p^a$ as $p^a\equiv v^a/\sum_av^a$, which obey the constraints $\sum_ap^a=(\sum_ap^a)^1=1$.\\
When the potential is taken into account, the complete behavior close to the singularity is given by a piecewise succession of Kasner solutions 'glued' together at the bounces against the billiard walls, where the role of the Kasner exponents is permuted, according to the BKL map as a function of the variable $u_{\rm BKL}$.\\
Decomposing the variables $\beta^a$ as $\beta^a\equiv\rho\gamma^a$, such that $\gamma_a\gamma^a=-1$, solving the Hamiltonian constraint for $\ln\rho$ is equivalent to project the dynamics on the unit hyperboloid $\gamma_a\gamma^a=-1$.\\
By means of appropriate geometrical transformations (for which no information about the dynamics is lost), it is possible to describe the dynamics on the UPHP \cite{yellow} \cite{Balazs:1986uj} \cite{terras}. In the UPHP, with coordinates ($u,v$), geodesics are (generalized) circles (with radius $r$) orthogonal to the $u$ axis (i.e. centered at ($u=u_0, v=0$)). They can be parametrized with their oriented endpoints on the $u$ axis, say $u^+$ and $u^-$, such that $r=\mid u^+-u^-\mid/2$ and $u_0=(u^++u^-)/2$. Although the complete phase space for billiards on the UPHP is in principle $4$-dimensional, the existence \cite{corn1982} of a reduced $2$-dimensional form $\omega(u^-,u^+)$,
\begin{equation}\label{omega}
\omega\equiv\frac{du^+\wedge du^-}{(u^+-u^-)^2},
\end{equation}
invariant under the billiard dynamics in the reduced $2$-dimensional phase-space $(u^-,u^+)$ explains the choice of the parametrization of geodesics according to their oriented endpoints as completely defining the billiard dynamics \cite{Damour:2010sz}, where the reduced phase space technique has been developped in Section IV and in the Appendix. Furthermore, the geometrical definition of probabilities, given by the integration of the form (\ref{omega}) over the pertinent subdomain of the $(u^-,u^+)$ reduced phase space, exactly coincides with the definition of probabilities for the statistical maps of the billiard, given in \cite{Chernoff:1983zz}, \cite{sinai83}, \cite{sinai85}.\\
After the projection on the unit hyperboloid, the expressions of the gravitational walls in term of the $\beta$ variables are equivalent to those in term of the $\gamma$ variables. The Kasner parameters $p^a$ can be expressed as a function of the $\gamma$ variables, and, on their turn, in terms of the Kasner scale factors $a$, $b$, $c$. The names of the three curvature walls in (\ref{curv}) below defining the three sides of the big billiard are due to the fact that the expression of the $\gamma$ variables as functions of the coordinates $u$ and $v$ of the UPHP reduce to that of the Kasner parameters $p$ as a function of the $u_{\rm BKL}$ variable at $u\equiv u^+$ and $v=0$. The role of the reflections on the curvature walls is then interpreted as the permutation of the Kasner parameters $p$ via the BKL map on the variable $u_{\rm BKL}$ (rigorously) identified with $u^+$. The role of the Kasner parameters will then be discussed according to all these relations.\\
\\
The BKL symmetry quotienting mechanism for the big billiard, as well as the CB-LKSKS ones, both for the one variable map for the BKL variable $u$ and for the two-variable map for the two quantities $u^-$ and $u^+$, encode the  sequence of Kasner epochs and Kasner ares which characterize the dynamics, and are based, on the one hand, on the statistical properties of billiard system, and, on the other hand, on the  continued fraction decomposition of thee oriented endpoints of the geodesics which define trajectories on the UPHP.\\
Furthermore, all the initial conditions for which the continued-fraction decomposition of the initial values of $u^+$ and $u^-$ for a given trajectory is limited  are to be excluded \cite{bel2009}.\\
Within this framework, the relevance of the initial values of the oriented endpoints has not been examined yet from the quantum point of view. The possibility to examine quantum-mechanical invariants containing information about the continued-fraction decomposition of the oriented endpoints offers a useful way to understand the relevance of the classical initial data problem from the quantum point of view.\\
From the quantum point of view, an aim of the present work will be the possibility to gain information on the wave function as far as these features are concerned.
\\        
%%%%%%%%%%%%%%%%%%%%%%%%%%%%%%%%%%%%%%%%%%%%%%%%%%%%%%%%%%%  
\subsection{The big billiard} 
\begin{figure*}[htbp]
\begin{center}
\includegraphics[width=0.7\textwidth]{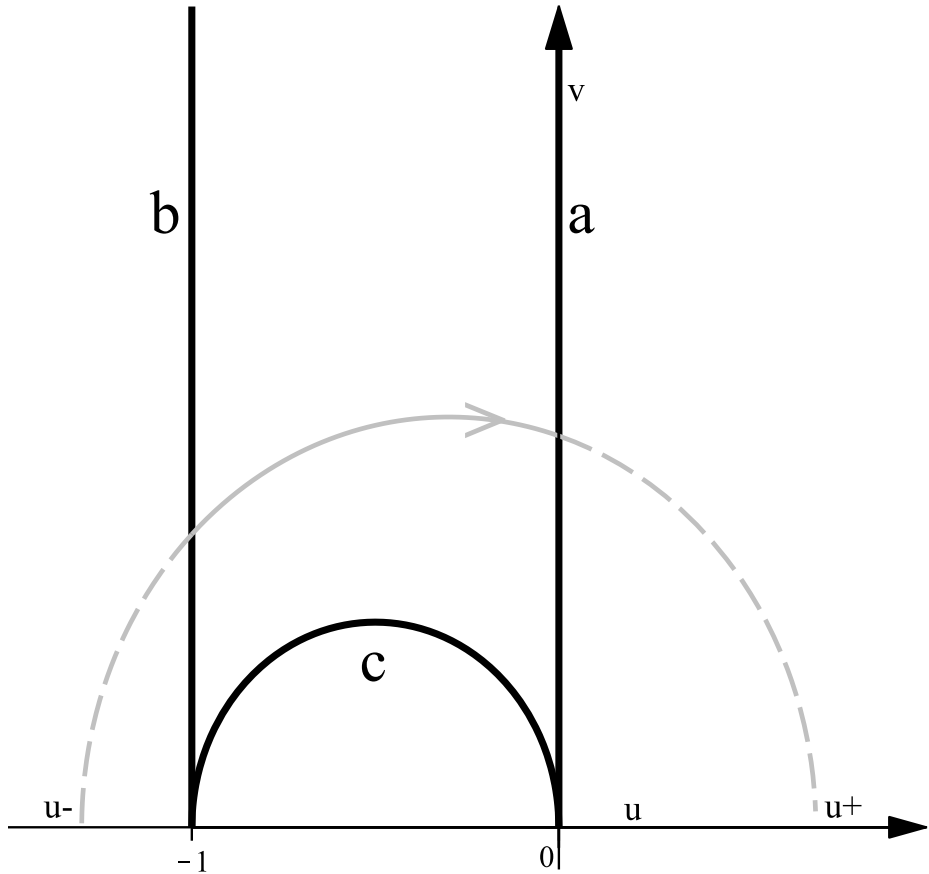}
\caption{\label{figura1} The big billiard table on the PUHP plane is sketched (solid black line). A $ba$ epoch, i.e. the oriented solid gray line starting from the side $b$ and ending on the side $a$, is parametrized by the oriented endpoints $u^-$ and $u^+$ on the $u$ axis of the corresponding geodesics (gray dashed line). }
\end{center}
\end{figure*}

\begin{figure*}[htbp]
\begin{center}
\includegraphics[width=0.7\textwidth]{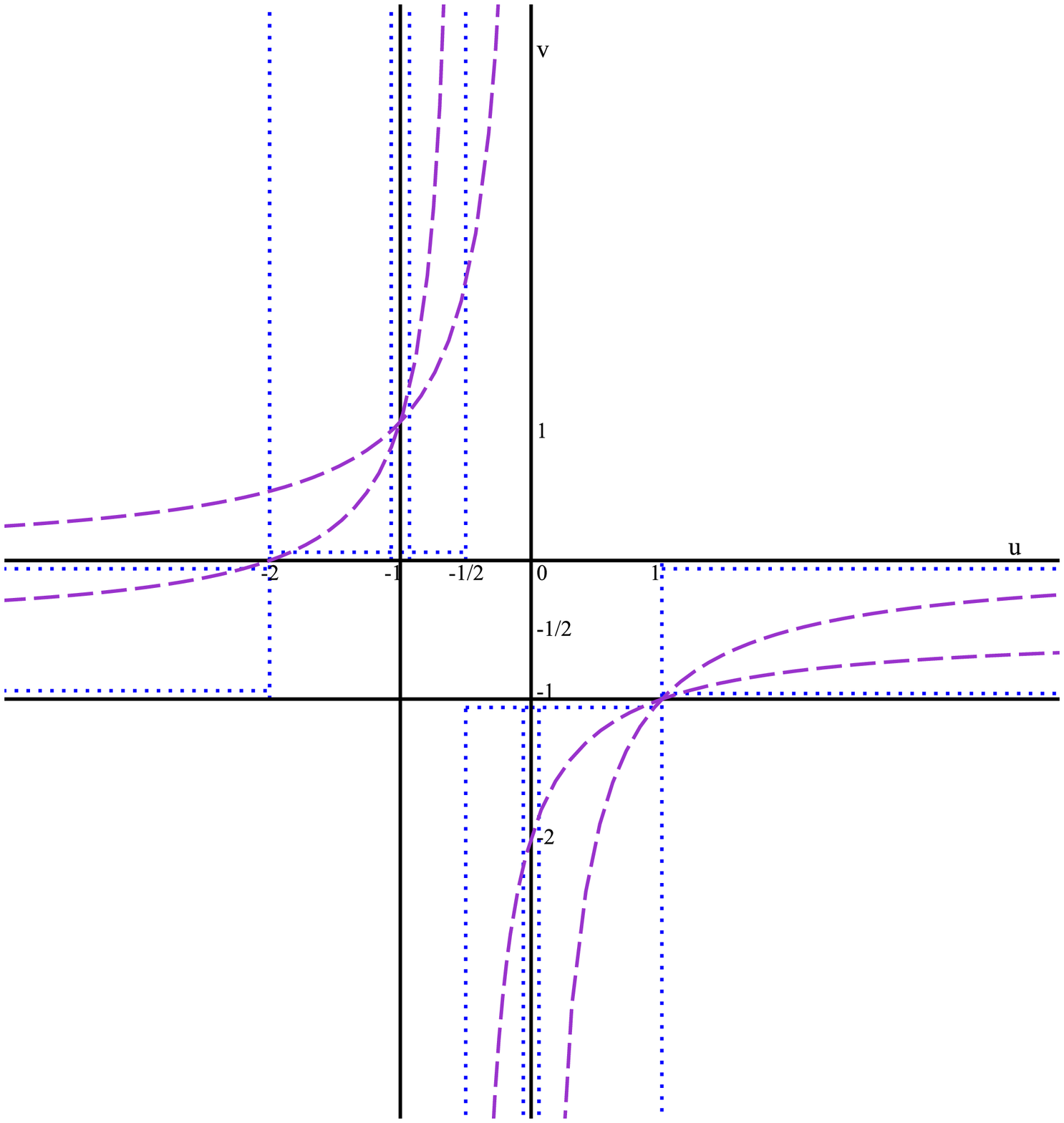}
\caption{\label{figura3} The different regions of the ($u^-u^+$) plane considered in the different billiard maps. The (thick solid) black lines delimit the boxes $B_{xy}$ for the epoch hopscotch of the big billiard. The (dotted) blue lines delimit the starting boxes $F_{xy}$ of the era hopscotch for the big billiard,  their details are given in \cite{Damour:2010sz}. The (dashed) purple lines delimit the domain where the dynamics of the small billiard takes place.}
\end{center}
\end{figure*}

In the UPHP, the big billiard consists of the billiard table delimited by the curvature walls $a$, $b$, $c$, and are described by reflections laws $A$, $B$, $C$ given by
\begin{subequations}\label{curv}
\begin{align}
&a:\ \ u=0, \ \ \ \ Au=-u\label{a}\\
&b:\ \ u=-1, \ \ \ \ Bu=-u-2\\
&c:\ \ \left(u+\tfrac{1}{2}\right)^2+v^2=\tfrac{1}{4}, \ \ \ Cu=-\tfrac{u}{2u+1},
\end{align}
\end{subequations}
where the transformations $A$, $B$ and $C$ act diagonally on $u^{\pm}$ (i.e. independently on $u^+$ and $u^-$). The big billiard table is sketched in Figure \ref{figura1}.\\
%%%%%%%%%%%%%%%%%%%%%%%%%%%%%%%%%%
%%%%%%%%%%%%%%%%%%%%%%%%%%%%%%%%%%%%%%%%%%%%%%%%%%%%
The billiard dynamics can be therefore parametrized according to: $i$) the 'free-flight' evolution between to consecutive reflections on the billiard walls, and $ii$) the three reflections $A$, $B$, $C$ against the three billiard walls $a$, $b$ and $c$.\\
The (appropriate composition of) the transformations $A$, $B$, $C$ define the big-billiard map $\mathcal{T}$, and the reduced form $\omega$ in Eq. (\ref{omega}) is invariant under the map $\mathcal{T}$. The billiard dynamics is therefore characterized by the bounces on the billiard walls, and the free-flight evolution is neglected. As a result, a Poincar\'e return map on the billiard walls is obtained.\\
The dynamics can be encoded as a sequence of epochs and eras. An epoch is defined as a trajectory (free-flight evolution) between any two walls, while an era is defined as a collection on epochs taking place in the same corner.\\
Accordingly, $6$ kinds of (oriented) epochs are distinguished, according to the walls connected by the corresponding trajectory. In more detail, these are $ab$, $ba$, $ac$, $ca$, $bc$, $cb$. Correspondingly, $6$ kinds of eras are obtained, and are named after the first epoch starting the era.\\
The reduced phase-space $(u^-u^+)$ can therefore be divided into $6$ different regions, where the $6$ kinds of epochs take place. These 'boxes' are named $B_{xy}$ after the corresponding $xy$ epochs; their details are given in Table II of \cite{Damour:2010sz}, and they are sketched in Figure \ref{figura3}, where the $B_{xy}$ boxes are delimited by solid black lines. The succession of epochs is described in this plane as a succession of 'jumps' in a epoch-hopscotch game. Probabilities for sequences of epochs and eras are defined by the integration of the reduced form $\omega$ in Eq. (\ref{omega}) over the pertinent regions of the $(u^-u^+)$ plane.\\
%%%%%%%%%%%%%%%%%%%%%%%%%%%%%%%%
%%%%%%%%%%%%%%%%%%%%%%%%%%%%%%%%
During a Kasner $xy$ era, the two Kasner exponents $x$ and $y$ oscillate (thus defining epochs), while the third one, say $z$, undergoes a monotonic evolution. In the next Kasner era, the $z$ exponent takes the place of one of the two previously-oscillating $x$ or $y$ (according to whether the previous era contains an odd or an even number of epochs).\\ 
\\
In addition, the compositions of the big-billiard map $\mathcal{T}$ allow one to describe the billiard dynamics of epochs, for which only the initial values $u^\pm_F$ determine both the type and the length (expressed as the number of epochs), and, according to them, also the following sequence of eras. The corresponding starting regions for $u^\pm_F$ in the $(u^-u^+)$ plane can be defined, and the dynamics is described as a succession of 'jumps' in an era-hopscotch game. The definitions of the $F_{xy}$ regions are given in Table III of \cite{Damour:2010sz}, and they are sketched in Figure \ref{figura3}, where the $F_{xy}$ boxes are delimited by (dotted) blue lines.\\
\\   
\subsection{The Kasner-quotiented big billiard}
The Einstein equations from which the billiard dynamics is derived are invariant under the group of permutations between the three functions $a$, $b$, and $c$. The corresponding billiard dynamics can be therefore quotiented by use of this group of permutations by defining BKL epochs and BKL eras.\\
During a single Kasner era, the role of the oscillating $x$ and $y$ exponents is identified in a BKL epoch, while that of the monotonically-evolving $z$ is kept fixed. The transition to the following BKL era is obtained by identifying the role of the $z$ exponent with either $x$ or $y$, and keeping the remaining one (either $y$ or $x$, respectively) as fixed.\\
This is equivalent to mapping all the starting boxes $B_{xy}$ (with the corresponding $F_{xy}$) onto a (preferred) one, say $ba$ (for technical purposes). This is obtained by means of the Kasner transformations $k_i$. The explicit expressions of the Kasner transformations $k_i$'s for the variables $u^+$ and $u^-$ are listed in \cite{Damour:2010sz}, Table I and Table VI. The Kasner transformations for the UPHP are listed in Table \ref{table1}; those for the restricted phase space are obtained by restricting the transformations of Table \ref{table1} on the absolute of the UPHP, at $v=0$; their geometrical interpretation in permuting the Kasner exponents, as well as their specific action with respect to the (preferred) epoch-type $ba$, are given in Table \ref{table1} as well.\\  
The corresponding quotiented map $T_{\rm BKL}$ is a succession of compositions of the Kasner transformations $k_i$ and the big-billiard map $\mathcal{T}$. The usual BKL map for the $u\equiv u^\pm\equiv u_{\rm BKL}$ variable, $T_{\rm BKL}$
\begin{equation}\label{BKL}
u\rightarrow u-1\rightarrow u-2 \rightarrow ... \rightarrow u-n-1 \rightarrow u'\equiv\tfrac{1}{u-[u]}-1,
\end{equation}
is then obtained. In Eq. (\ref{BKL}), $[u]$ denotes the integer part of $u$, and $[u^+]=n-1$, $n$ being the number of epochs in the $u$ era, such that $0\le[u^+_F]\le n-1$. The $BKL_{u>0}$ definition for the $u_{\rm BKL}$ variable, defined in Eq. (5.4) of \cite{BLK1971} and used in \cite{Damour:2010sz}, is used here in Eq. (\ref{BKL}); as discussed in \cite{Damour:2010sz}, this definition is more suitable to naturally describe the features of the billiard. Of course, also the $BKL_{u>1}$ definition of \cite{llft} can be used, after a proper relabelling (rescaling) of the variables.\\ 
In Eq. (\ref{BKL}), it is possible to identify two different features: $i$) the oscillatory regime between two Kasner exponents (say $a$ and $b$) in the part $u\rightarrow u-1\rightarrow...$, with the other Kasner exponent (say $c$) evolving monotonically, i.e. the 'BKL epoch map'; and $ii$) the transition to the next BKL era, when the $c$ exponent takes the place of one of the two previously-oscillating exponents (whose role is quotiented in the BKL map), i.e. the 'BKL era-transition map', corresponding to the $u'\equiv\tfrac{1}{u-[u^+]}-1$ element.\\
In \cite{Damour:2010sz}, the BKL epoch map is given in Table IV, while the corresponding subregions of the restricted phase space are sketched in Figure 6; the BKL era map is given in Table V, and the pertinent subregions of the restricted phase space are depicted in Figure 7.\\
During a BKL era, the BKL epochs correspond to geodesics with the same radius $r$, and cantered at $v=0$ with $u_0$ scaling according to Eq. (\ref{BKL}). The next BKL era consists of geodesics of different radius $r'$ and different center $u_0'$, $v=0$, $r'$ and $u_0'$ being defined by the BKL era-transition map.\\ 
\\
It is possible to extract from the BKL map (\ref{BKL}) information about the first epoch of each era only, and neglect the succession of epochs in each era, considering the boxes $F_{xy}$ only. For the particular choice of the $ba$-epoch type, the era-hopscotch dynamics is the Chernoff- Barrow- Lifshitz- Khalatnikov- Sinai- Khanin- Shchur map $\mathbf{T}$ for the big billiard, or CB-LKSKS map in short. The two-variable CB-LKSKS map was given in \cite{Chernoff:1983zz}, in the first part of Eq. (4) (with different definitions for the variables with respect to the notation used elsewhere) as a four-variable map, in the complete Eq. (4), and in \cite{sinai83}, Eq. (4), and in \cite{sinai85}, Section 4, for the two-variable unit-square map, which encodes the properties of the fractional parts of the two-variable BKL map. This map is defined for the variables $u^+$ and $u^-$ in Eq. (6.2) of \cite{Damour:2010sz} as
\begin{equation}\label{cblksks}
{\bf T} u_F^\pm=+\frac{1}{u_F^\pm-[u_F^+]}-1.
\end{equation}
In this case, the properties of the dynamics are encoded in the succession of eras, defined by their first epoch only. In particular, the CB-LKSKS map is a succession of BKL era-transition maps. Correspondingly, a Poincar\'e return map on the $a$ wall for the first epoch of each quotiented $ba$ era is obtained.\\
Both the BKL map $T_{\rm BKL}$ and the CB-LKSKS map are based on the identification of the role of different Kasner exponents, and therefore destroy the typical 'bouncing' description of billiards, based on reflections on the billiard sides. As a result, a BKL era $ba$ is described by a succession of translated $ba$ epochs, where no bounces are considered.\\ 
One of the key-features of these maps is the possibility to extract the exact number of epochs contained in the succession of eras from the continued-fraction decomposition of the variable $u^+$; this feature has allowed for a rigorous statistical analysis of the BKL dynamics. In particular, this expression has a very simple functional expression on the case of $ba$ epochs and eras. Nevertheless, it is possible in principle to define a BKL map and a CB-LKSKS map for each one of the $xy$ kinds of epochs and eras; their explicit form would be more complicated than Eq. (\ref{BKL}), and the extraction of the number of epochs in each era less immediate.\\
\\
%%%%%%%%%%%%%%%%%%%%%%%%%%%%%%%%%%%%%%%%%
\subsection{The small billiard}
\begin{figure*}[htbp]
\begin{center}
\includegraphics[width=0.7\textwidth]{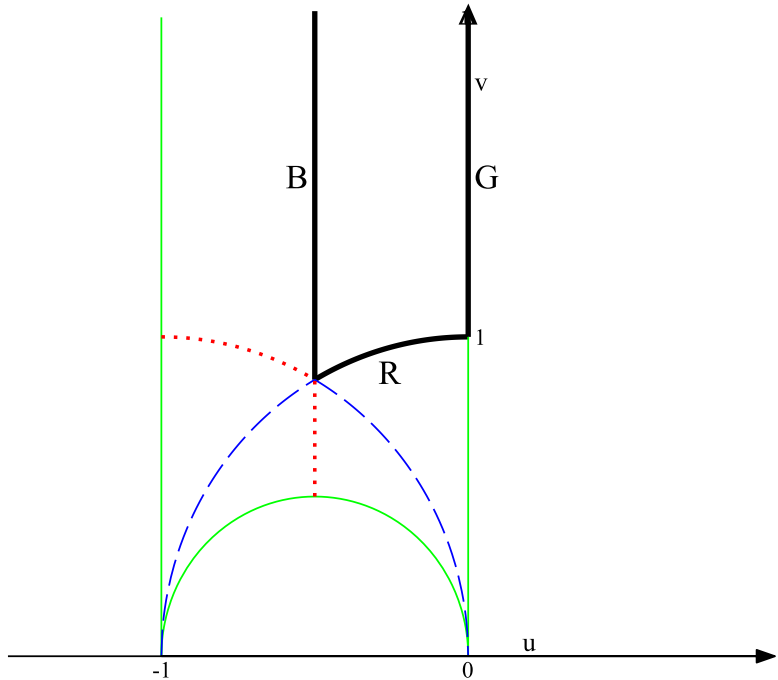}
\caption{\label{figura2} The small billiard table on the Poincar\'e upper half plane is sketched (thick solid black line).\newline
The 'symmetry lines' of the big billiard (corresponding to the other subdominant symmetry walls for the inhomogeneous case) are also displayed; in particular, the blue (dashed) lines are the part of the symmetry walls that bisects a given corner, while the red (dotted) lines are the part of the symmetry walls perpendicular to the side opposite to the given corner. The green (solid thin) lines are the sides of the big billiard (which correspond to the other curvature walls).\newline
The colors (dotted lines, dashed lines and thin solid lines) of the symmetry lines of the billiard describe how to 'glue' the $6$ equivalent copies of the small-billiard table and obtain the big-billiard table in Figure \ref{figura1}. In more detail, the $G$(reen) side has to coincide with the green (solid thin) line, the $B$(lue) side has to coincide with the blue (dashed) line, and the $R$(ed) side has to coincide with the red (dotted) line.}
\end{center}
\end{figure*}

\begin{figure*}[htbp]
\begin{center}
\includegraphics[width=0.7\textwidth]{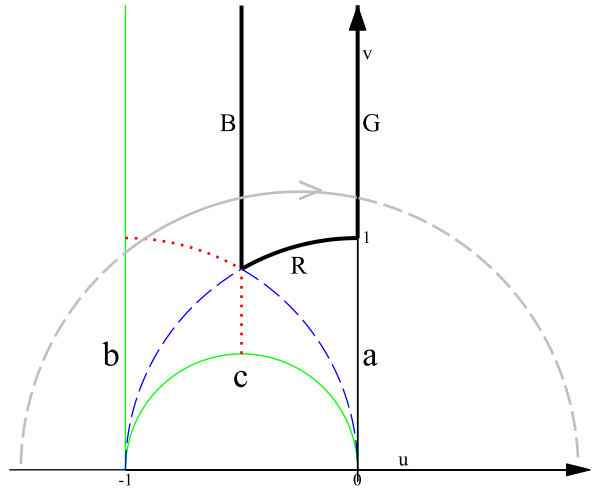}
\caption{\label{figura1new} Integrated description of the UPHP, which shows the big billiard table, the small billiard table, the symmetry lines of the big billiard wich define the small billiard, and an oriented geodesics that defines a trajectoty in the billiard.}
\end{center}
\end{figure*}

The small billiard is obtained in the more general non-homogeneous case.\\
The small-billiard table is delimited by the three walls $G$, $B$, $R$, which stand for $G$reen, $B$lue and $R$ed, respectively, and whose corresponding reflection laws define the small-billiard $t$ map
\begin{subequations}\label{t}
\begin{align}
&G: \ \ u=0, \ \ \ \ Gu= -u,\\
&B: \ \ u=-\tfrac{1}{2},\ \ \ \ Bu= -u-1\\
&R: \ \ u^2+v^2=1, \ \ \ \ Ru= \tfrac{1}{u}.
\end{align}
\end{subequations}
In particular, The $G$ wall consists of part of the curvature wall $a$ in Eq. (\ref{a}), while the $B$ wall and the $R$ wall are (parts of) symmetry walls. The 'color' names $G$, $B$ and $R$ are guessed from \cite{Damour:2010sz}, where the restricted phase space for the small billiard is established in Table VIII and represented in Figure 8; the unquotiented small billiard map for the variables $u^+$ and $u^-$ is defined in Eq.'s (9.3), (9.4) and (9.5); the possible symmetry-quotienting mechanisms for the small billiard are discussed in Subsection IX B.\\
The small billiard table is sketched in Figure \ref{figura2}, where a possible use of the 'color' names is depicted: different plotstyles mimic the role of colors.\\
%%%%%%%%%%%%%%%%%%%%%%%%%%%%%%%%%%%%%%%%%%%
%%%%%%%%%%%%%%%%%%%%%%%%%%%%%%%%%%%%%%%%%%%%%%%%%%
Epochs can be defined also in this case as trajectories joining any two walls, but, according to the geometry of the small-billiard table, small-billiard eras are defined as a sequence of epochs between two reflections on the $R$ side. The regions of the reduced phase-space ($u^-,u^+$) where the dynamics of the small billiard takes place are described are
\begin{subequations}\label{table4}
\begin{align}
&BG : u^-<-1,\ \ u^+>u_\alpha;\ \  -1<u^-<-1/2,\ \ u^+>u_\beta;\\
&BR :   u^-<-1,\ \ u_\beta<u^+<u_\alpha;\\  
&RG :   -1<u^-<-1/2,\ \ u_\alpha<u^+<u_\beta;\ \ -1/2<u^-<0,\ \ u^+>u_\alpha;\\ 
&RB :   -1/2<u^-<0,\ \ u^+<u_\beta;\ \ \ \ 0<u^-<1,\ \ u_\alpha<u^+<u_\beta;\\ 
&GR :   u^->1,\ \ u_\alpha<u^+<u_\beta;\\  
&GB :  0<u^-<1,\ \ u^+<u_\alpha;\ \ u^->1,\ \ u^+<u_\beta,
\end{align}
\end{subequations}
 where the two functions $u_\alpha$ and $u_\beta$ are defined as $u_{\alpha} \equiv - \frac{1}{u^-}$ and $u_{\beta} \equiv - \frac{u^- + 2}{2u^- + 1}$, respectively. The regions of the $(u^-,u^+)$ allowed for the dynamics of the small billiard are classified according to reflection laws on the boundaries of the small billiard. In particular, the ($u^-,u^+$) plane is divided into $6$ subregions allowed for the oriented endpoints of trajectories joining the three different walls $G$, $B$ and $R$, and they are sketched in Figure \ref{figura3}, where the subregions are delimited by (dashed) purple lines.
The small-billiard is then efficient in reproducing the dynamics of the unquotiented big billiard in the following sense.\\
The symmetry walls that define the six 'small billiards' of which the big billiard can be constructed of can be considered as 'symmetry lines' of the big billiard itself, as sketched in Figure \ref{figura2}. In particular, the epochs of each era of the big billiard always cross the $B$ line, while the crossing of the $R$ lines cannot be defined \textit{a priori}, as one can verify directly. This behavior is in one-to-one correspondence with the dynamics of the small billiard, which shows when one of the the two oscillating factors vanishes (the side $G$ of the small billiard is hit, and one of the sides of the big billiard is hit), when they equal each other (the side $B$ of the small billiard is hit, and a $B$ line is crossed by a trajectory of the big billiard), and when the non-oscillating Kasner exponent crosses one of the other two when changing its slope (the $R$ side of the small billiard is hit, and the trajectory crosses a $R$ line in the big billiard). Consequently, each era of a given type (say $ba$)
 of the unquotiented big billiard can be further (sub)-classified according to whether the first epoch and the last epoch of the era cross the $R$ line or not. The sub-classification then consists of five different $R$-crossing properties for each era.\\
The comparison of the dynamics of the small billiard as far as maps are concerned can be obtained by considering the Poincar\'e return map on the $G$ wall. The region of the $(u^-u^+)$ plane where the dynamics take place shows that small-billiard eras contain both $ba$ and $ca$ big-billiard epochs in the $RG$ and the $BG$ regions. In particular, the small billiard dynamics reconstructs the complete $R$-crossing repertoire for an era, but selects some of them from the $B_{ba}$ box, and others from the $B_{ca}$. As a result, the small billiard fails in reproducing the BKL map because BKL eras are defined according to the Poincar\'e return map on the billiard walls, while small-billiard eras according to the hit on the $R$ wall. The full picture involving the big billiard table, the big billiard table and an oriented geodesics defining a trajectory on the UPHP are sketched in Figure \ref{figura1new}.\\
\\
Nevertheless, a CB-LKSKS map will be constructed for the small billiard, and the meaning of a $n$-epoch BKL era can be recovered.\\
\\
%%%%%%%%%%%%%%%%%%%%%%%%%%%%%%%%%%%%%%%%%%%%%%%%%%%%%%%
%%%%%%%%%%%%%%%%%%%%%%%%%%%%%%%%%%%%%%%%%%%%%%%%%%%%%%%%%
\subsection{Several Descriptions of the cosmological singularity}
The BKL approach to the cosmological singularity is an extremely powerful tool to 
investigate the behavior of the early universe in this asymptotic limit.\\
In the literature, many studies have been presented, whose aim is both to explore the
implications of the BKL conjecture from different points of view, and to try to
add back (some of the) features, which, in the BKL approach, are considered as negligible in the asymptotic limit toward the cosmological singularity, such as
the presence of spacial derivatives.\\
Furthermore, the choice to use Iwasawa variables is based on their effectiveness in
solving the Hamiltonian constraint via a straightforward geometrical projection
rather than considering the square root of the corresponding quantity. Moreover,
the geometrical interpretation of the Iwasawa variables allows one to gain information about the geometrical structure of the billiard table(s) obtained in the asymptotic limit.\\
Adopting the Iwasawa decomposition of the metric tensor allows one to outline the symmetry properties of the ordinary $4=3+1$ dimensional description, which is found to account for the proper limit of higher-dimensional models [T. Damour, S. de Buyl, M. Henneaux 1 and C. Schomblond, J. High Energy Phys. 0208 (2002) 030, arXiv:hep-th/0206125], [T. Damour, M. Henneaux and H. Nicolai, Phys. Rev. Lett. 89 (2002) 221601, arXiv:hep-th/0207267]
[T. Damour, M. Henneaux, B. Julia and H. Nicolai, Phys. Lett. B 509 (2001) 323, arXiv:hep-th/0103094], [T. Damour and M. Henneaux, Phys. Rev. Lett. 86 (2001) 4749, arXiv:hep-th/0012172].\\ 
The billiard interpretation of higher dimensional models, where the billiard walls are defined according a different paradigm, is discussed in \cite{Fre':2005si}, and the succession of the corresponding sequence of free-flight piece-wise approximations, which are interpreted as Kasner epochs as far as the evolution of the present universe is concerned, is presented in \cite{Fre:2005bs}\\
\\
The Misner variables and the Misner-Chitr\'e variables offer a direct physical interpretation \cite{Misner:1969hg}, \cite{chi1972}, \cite{Misner:1994ge}, \cite{Misner:1974qy}: a variable can be isolated, which accounts for the isotropic growth of the universe and can therefore be adopted as a suitable Hamiltonian time variable, while the remaining variables are interpreted as the anisotropic degrees of freedom in the universe evolution, and can be treated as space variables in the Hamiltonian scheme. Nevertheless, it is possible to demonstrate that the asymptotic limit towards the cosmological singularity (i.e. the billiard representation of the UPHP) is exactly the same for both models.\\  
\\
The so-called Iwasawa conformal Hubble-normalized orthonormal frame variables are another set of variables used in investigations in quantum cosmology. In the dynamical-system approach, the solution of the Einstein equations is described by a state space associated to a state vector composed of the diagonal components of the traceless shear matrix, the Fermi rotation variables, and the spatial commutation functions (i.e. the connections) that describe the $3$-curvature associated to the conformal metric.\\
Within this scheme \cite{wain} \cite{uggla2003} \cite{uggla2013}, the implementation of the BKL asymptotic dynamics yields a description of the Kasner circle which is to some extent 'dual' to that resulting from the Misner picture and from the Iwasawa framework. In fact, within this description, straight lines are obtained in the Kasner circle at each change of epoch, while points on the Kasner circles are obtained for the regimes of the solution to the Einstein equations, for which the derivative of the Kasner trajectories with respect to the parametric time is constant, i.e. during the free-flight evolution.\\
In \cite{ringstrom2000}, an analysis has been produced of curvature scalars and causal geodesics for the Bianchi IX asymptotic description, according to different initial data set. As a result, the model has been proved to have curvature singularities in the limit close to the cosmological singularity.\\
Within the same variables, the asymptotic behavior of Bianchi models has been analyzed for different kinds of matter (i.e. if inhomogeneities are taken into account) included in the model \cite{Ringstrom:2000mk}. The properties of different kinds of matter, according to their different characterization in the right-hand side of Einstein equations, have been investigated as far as their capability to modify the oscillating features of the solution of the Einstein equations is concerned. As a result, the chaotic behavior has been found for matter whose stress-energy tensor is different from that of a stiff fluid.\\
\\
The existence of the so-called Mixmaster dynamics, i.e. the features for the asymptotic limit of the Bianchi IX universe for which the quantities assimilated to the anisotropic scale factors (the generalized Kasner parameters of each piece-wise solution of the Einstein equations) have opposite behavior and give rise to Kasner epochs and Kasner eras, has been tested by means of several investigations.\\
\\
As far as the Bianchi classification of the solution of the Einstein equation is concerned, the asymptotic solution of the Bianchi IX model interpreted as a sequence of geodesic evolution followed by a bounce on the sides of the billiard table can be locally referred to as a sequence of Bianchi I and Bianchi II solutions, respectively.\\
\\
In \cite{Heinzle:2009eh}, the asymptotic behavior towards the cosmological singularity for the Bianchi IX universes has been proved to consist of sequences of Bianchi I and Bianchi II piecewise solution, corresponding to geodesic free-flight evolution and bounces against the billiard table, and that there exist a set of initial conditions such that the solution is of the kind of Taub (i.e. the billiard ball reaches the cosmological singularity). Within this analysis, the discrete permutation symmetries among the sides of the billiard play a crucial role, such that the behavior of different Bianchi models is predicted to have a less symmetric description.\\
\\
In \cite{Berger:2000hb}, the Mixmaster properties of the dynamics have been tested by assuming a line element whose geometrical features are simpler than those of the Bianchi IX model, and the presence of epochs and eras has been looked for. In this work, a $U(1)$ symmetric cosmology on $S^3\times R$ has been taken into account: this model can be considered as geometrically simplified with respect to the usual Bianchi IX model, as there is one Killing vector less. As a result, a  behavior which can be reconstructed to the presence of Kasner epochs and Kasner eras has been verified also within this scheme, as a further confirmation that space points can be considered as spatially decoupled in the asymptotic limit to the cosmological singularity \cite{montannphys}.\\
More in detail, the explicit Killing direction for the $U(1)$ symmetric models is chosen as one of the $3$ space directions, such that the square of the norm of the Killing field equals the corresponding element of the Bianchi IX metric tensor. The behavior of the logarithm of the norm of the Killing field is demonstrated to exactly overlap that of the logarithmic scale factor of smallest absolute value of the Bianchi IX model with respect to the parametric time, and so its derivative, both for epochs and era changes. In fact, this variable is shown to be approximated, under the corresponding conditions, i.e. at each epoch change, to one of the three Bianchi IX logarithmic scale factors. The remaining variable that fully characterizes the spatial symmetries of the model, furthermore, is shown to monotonically decrease (with respect to the parametric time evolution) and to exhibit the opposite behavior only for the era change. These features are 
this way interpreted as 
verifying the BKL paradigm also in the inhomogeneous case, i.e. the hypotheses according to which points are spatially decoupled in the asymptotic limit close to the cosmological singularity are tested also within the framework of the different spatial metric adopted in this work.\\
\\
A very accurate numerical simulation of a generic spacetime in vacuum without any particular symmetry was presented in \cite{garf93}. Because of the generality of the model analyzed, the result are interpreted as describing the general behavior of singularities: this general behavior is found to be local (i.e. describing spatially decoupled points) and oscillatory. Furthermore, the behavior of the early universe in absence of spatial derivatives and that in presence of spatial derivatives have been contrasted. The differences in the behavior have been attributed to the effects of spatial derivatives on the eigenvalues of the matrix that generalizes the shift vector in the spectral method used in the numerical simulation. More in detail, both in absence of spatial derivatives and in presence of spatial derivatives the sequence of bounces (i.e. the sequence of BKL epochs and eras) is the same, the only difference being that, in presence of spatial derivatives, this sequence appears more rapidly, such that
the effects of the presence of spatial derivatives in the general solution of the Einstein equations has been a challenging investigation for the last years.\\
\\
The appearance of spikes is a striking feature of the early universe in this more general characterization.\\
Within this framework, the Mixmaster behavior, which is modified by these spatial structures, is explained as a part of a more complicated scheme, which admits a greater set of (not yet analyzed) symmetries \cite{uggla2012}. In particular, 'spikes' are described as the result of a piecewise sequence of solutions, locally describing a sequence of Bianchi I and inhomogeneous vacuum models.\\
From the geometrical point of view, spikes are described as spatial structures, whose main contribution to the modification of the BKL conjecture is to modify the curvature scalars according to the introduction of spatial gradients.\\
From the statistical point of view, spikes are described by the modification of the BKL epoch map, such that the $u$ variable undergoes the map $u\rightarrow u-2$. In the era-transition map, the presence of spikes is described by maps for the variables $u$, which do not follow the usual CB-LKSKS map. In particular, transformations of the kind $u\rightarrow 1/(u-2)$, $u\rightarrow -1+(u-1)^{-1}$, $u\rightarrow (-1+(u-1)^{-1})^{-1}$ are found, according to different intervals within which the variable $u$ is defined. These features accounts for curvature transitions which are not present in the usual BKL description.\\
The appearance of spikes is represented within the ('dual') description of the Kasner circle as trajectories which are different from those originated in the usual BKL paradigm.\\
An analysis of teh changes induced in the statistical properties of the BKL map by the presence of spikes has been performed in \cite{Bini:2008qg}, as far as the geometrical features of the Weyl scalar(s) are concerned, thus extending the Petrov classification for the Kasner solutions \cite{Cherubini:2004yi} to this more general asymptotic case.\\
\\
Higher-order spike transitions have been observed in \cite{wct}. These transitions are characterized by the map for the $u$ variable $u\rightarrow u-3$, and their behavior has been shown to decompose as two first-order spike transitions, i.e. for the maps $u\rightarrow u-2$ and $u\rightarrow u-1$.\\
The occurrence of spikes has also been recently analyzed within the framework of the physical interpretation of the results found with the so-called 'solution-generating techniques'. Some solution-generating techniques \cite{uggla2013} have the same effect of performing a rotation (with respect to the Fermi non-propagating frame of the state vector of these models). Two kinds of approaches can be followed, namely considering the asymptotic behavior or considering their explicit expressions. In the second case \cite{uggla2013}, the explicit solution turns out to be a one-parameter family of Kasner solutions.\\
\\
The features connecting the particular maps of the unquotiented small billiard, the effect of the presence of space derivatives in 'speeding up' the succession of Kasner trajectories and the occurrence of spike transitions on the case of spatial gradients are phenomena which would deserve a more detailed comparison, as they might seem to have a closer correlation.\\
\\
The features of chaos which characterizes the BKL solution can be statistically analyzed by evaluating the Lyapunov exponents for the BKL map. In the case of a two-variable map, the variable $u^+$ is found to have a positive Lyapunov exponent, while the variable $u^-$ is found to have a negative Lyapunov exponent. As a result, the variable $u^+$ is described as unstable under the BKL map, while the variable $u^-$ is found to be stable under the BKL map. More in detail, these features can be accounted for by recalling that the variable $u^+$ encodes the 'future' evolution for the billiard representation of the solution of the Einstein equations, while the variable $u^-$ 'stores' the past evolution of the billiard motion. In other words, a small perturbation of the initial value of $u^+$ can be shown to result in a modification of the future evolution of the billiard behavior, which is totally impredictable (i.e., chaotic), as the quotiented BKL variable $u^+$ is defined within the range $0\le u^+\le\infty$, 
while a modification of the initial value of $u^-$ does not imply any change in the BKL sequence of epoch and eras, as the BKL map in the 'future' time direction is not determined by the value of $u^-$.\\
\\
Within this analysis, it is extremely useful to study the periodic configurations, i.e. the initial values of the variables $u^+$ and $u^-$ according to which a periodic sequence of epochs and eras is obtained.\\
In the BKL quotiented big-billiard description, a periodic orbit is represented by a continued-fraction decomposition of the $u^+$ variable, characterized by the periodic sequence of BKL eras in each epoch. In the unquotiented big-billiard description, the definition of periodicity is broader than that found in the quotiented version, and accounts also for the corner in which the oscillations take place, and for the orientation of each first epoch of each era. The features of the periodic phenomena of the chaotic motion inside the unit hyperboloid (instead of those projected on the unit hyperboloid) exhibit still a different characterization.\\
\\
Within this analysis, and, in particular, as far as the BKL quotiented big billiard is concerned, the occurrence of one-epoch periodic orbits is associated by points in the restricted phase space, corresponding to the so-called golden ratio and its conjugate, while the occurrence of $n$-epoch periodic eras is associated to points in the restricted phase space, corresponding to the so-called ($n$) silver means and their conjugates. The probability of a single $n$-epoch era to take place, as well as the probability for a finite sequence of eras to take place, is determined by a finite value, expressed by the integral of the corresponding region in the restricted phase space according to its non-trivial measure. On the contrary, the occurrence of an infinite sequence of eras, and, in particular, the occurrence of a periodic orbit, is represented by points in the restricted phase space, for which it is not possible to evaluate a probability within this framework.\\
Combining these two items of information together, i.e. considering the Lyapunov stability issues of the BKL map and the restricted phase space integral that allows one to express the probability for a particular sequence of orbits to take place, it is possible to define the regions of the restricted phase space within which the perturbation of an $n$ periodic orbit allows one for a finite sequence of $n$-epoch eras before the sequence is chaotically modified according to the instability of the $u^+$ variable under the BKL map. These regions of the restricted phase space can be shown to be related by suitable functions (ratios) of Fibonacci numbers and generalized Fibonacci numbers \cite{golden}, which depend on the the value $n$. The numerical analysis of \cite{berger1991} is crucial in proving that, despite the possibility to modify the definition a time variable, such that a vanishing Lyapunov exponent can be associated to the BKL  one-variable map, the proper time of the minisuperspace representation is 
the suitable time variable, according to which the BKL phenomena are described. Furthermore, the ranges of the BKL variable, where $1$-epoch periodic orbits are modified as a finite sequence of $1$-epoch eras, have been demonstrated to range between the intervals given by the suitable Fibonacci numbers, for the suitable time variable.\\
\\
From a broader perspective, within the same direction of the most recent investigations, the relations between the algebraic structures defining cosmological billiards and the Fibonacci numbers, as well as their more detailed characterizations and generalizations, has been investigated in \cite{carb}.
\\
The complete $4$-variable CB map \cite{Chernoff:1983zz} (named after Chernoff and Barrow) is obtained for the set of $4$ variables which accounts for the $3$ scale factors and a time variable, when the statistical map is evaluated on a suitable Poincar\'e section. Considering a Poincar\'e surface of section eliminates one degree of freedom in the solution of the Einstein equations, such that the CB map describes the oscillating features of the solution to the Einstein equations as a two-variable map for two variables (where the features of the BKL map can be traced), and two other variables, which account for overall scale factors. The complete CB $4$-dimensional map for the mixing behavior of the epoch BKL map and of the era BKL map has been numerically investigated in \cite{berger1993}. Within this framework, several fundamental aspects of the complete map have been related to the time-reversal features of the Einstein equations. In particular, the $4$-dimensional map has been numerically demonstrated to 
be 
effective in exactly reproducing the oscillation of two scale factors and the monotonic behavior of the remaining one during a BKL era by showing that, at each step of the BKL epoch map in a BKL era, (a suitable function of) the slope of the non-oscillating scale factor (described by the suitable time variable) remains constant. This properties has been tested by considering the complete $4$-dimensional CB map, which accounts also for a time variable, while the BKL maps (both the $1$-variable map and the $2$-variable map) do not include any time variable. As the $u^+$ variable encodes the sequence of 'future' epochs, while the $u^-$ variable accounts for the 'past' evolution of the billiard dynamics, it is possible to infer that reversing the order of the era sequence encoded in the continued-fraction decomposition of the $u^+$ variable corresponds to a time-reversal operation on the solution of the Einstein equations. The validity of this reasoning has been proven to be statistically correct, as the 
validity of the complete $4$-dimensional map has been investigated at extremely high numerical precision, where the complete map contains the time variable as well.\\
The relationship between the complete $4$-dimensional map and the BKL map has been numerically investigated in \cite{berger1994}. Within this framework, the complete $4$-dimensional map as been numerically proven to correspond to the biproduct of the usual BKL map and the map for the remaining variables, as theoretically stated in \cite{Chernoff:1983zz}.\\
The comparison of the complete $4$-dimensional map and the BKL oscillating mechanisms has been further verified in \cite{berger1997} by means of numerical simulation of long sequences of BKL eras. In particular, the $4$-dimensional map accounting for two oscillating scale factors, a non-oscillating scale factor and a time variable has been implemented, and the pertinent BKL behavior has been pointed out. This way, by means of very accurate numerical simulations, the effectiveness of the BKL map in quotienting the symmetries of the big billiard dynamics by properly exchanging the role of the Kasner variables has been further tested. Moreover, the relevance of the role of the big billiard in the description of the quotiented BKL dynamics has been pointed out also as far as numerical simulations are concerned.\\ 
\\
In most the investigations of the dynamics of the early universe within the framework of the BKL paradigm, much attention and strong efforts have been devoted to the study of the geometrical properties of the billiard table and of the trajectories followed by the billiard ball. In particular, the group theoretical structures that have been implemented up to now are based on the geometrical features of the billiard table, for which bounces on the billiard sides can be described by reflections.\\ 
\\
A thorough investigation and a rigorous explanation of group construction of the fundamental domain on the UPHP for cosmological billiards is given in \cite{Fleig:2011mu}, where, also in higher dimensions, the shape of the fundamental domain is provided on the higher dimensional UPHP, and an elegant formula for the volume is stated.\\
The relevance of this result is based on the possibility to generalize the present analysis to a higher number of space time dimensions.\\ 
\\
An easy solution of the Hamiltonian constraint is one of the advantages of the use of Iwasawa variables.\\
The full picture in which the Einstein field equations are mapped to is indeed that of the chaotic motion of a point particle inside the unit hyperboloid, which is coordinatized by the functions $\beta^\mu=\rho\gamma^\mu$. Solving the Hamiltonian constraint for the gravitational action with respect to the variable $\lambda=\ln\rho$ is equivalent to projecting the motion \textit{inside} the hyperboloid \textit{onto} the surface of the hyperboloid itself (i.e. one degree of freedom is eliminated). After the elimination of the Hamiltonian constraint, the projection on the unit disk and on the UPHP are considered only for the sake of a better visualization and of easier calculation.\\
\\
Some features of the quantum version of this model on the unit hyperboloid have already been performed in previous analyses.\\
In \cite{Kleinschmidt:2009cv}, the complete wave function (which contains also the variable $\rho$) is factorized, and, as a result, the complete wave function and all its $\rho$ derivatives tend to zero as the cosmological singularity is approached, when Neumann boundary conditions are taken into account. More in general, the full wavefunction is demonstrated to be in general complex and oscillating. In \cite{mk11}, a construction of wavepackets for quantum cosmological billiards by means of a suitable Gaussian weighting function has been proposed.\\ 
\\
If one chooses other kinds of variables, the square root of the Hamiltonian constraint has to be considered. When the momenta in the classical description have to be implemented to operators, a problem of factor ordering arises, i.e. when non-trivial geometries are considered, the differential operator acting on the wave function can be constituted by different items. Within this framework, and, in particular, within the framework of the quantum regime of the asymptotic limit of the Bianchi IX models, it is customary to choose a factor ordering in the Hamiltonian operator, for which the usual Laplace-Beltrami operator is reconstructed \cite{primordial}.\\ 
In this respect, in the work \cite{puzio}, the possibility to take into account the square root of the Laplace-Beltrami differential operator as a Hamiltonian, and to consider the second power of the eigenvalues of the corresponding (squared) eigenvalue equation, is analyzed; the conditions which have to be fulfilled are investigated. More in detail, the square root of the Laplace-Beltrami operator is interpreted to have the same eigenfunctions of the Laplace-Beltrami operator, but with corresponding eigenvalues being the second power of the eigenvalues of the original problem. Because of the features of this Hamiltonian operator, the time evolution of the wavepackets is different from that of the original ones, the operator being non-local. The second iteration of the operator leads to the Laplace-Beltrami operator itself, i.e. to the Laplacian for the UPHP, which can be identified with a Klein-Gordon operator. The only difference between this Klein-Gordon operator and the standard one is the fact that it 
is not possible to assign proper initial conditions for 
 the wave function and for its first derivative, as it would be for a second-order differential equation. The evolution of the wavefunction is analyzed by considering a simplified (with respect to the small billiard) domain, and the wavefunction is explained to be 'well-behaved' with respect to this simplification.\\
In more general settings, the Hamiltonian constraint assumes more complicated features, and other kinds of constraints have to be taken into account, when the quantum version of a cosmological model is considered. In particular, within the framework of supersymmetry and supergravity, further constraints appear. Supersymmetric quantum cosmology is a field which offers several directions of investigation; a complete analysis of the different problems can be found in \cite{monix}, where the constraints arising in superysmmetric quantum cosmology are thoroughly analyzed.\\
In \cite{Kleinschmidt:2009cv}, the supersymmetric version of cosmological billiard is presented. In particular, the Hamiltonian constraint is shown to be vanishing as a consequence of the supersymmetry constraint in the case of $11=10+1$ maximal supergravity.\\
In \cite{damnew}, the specific case of $4=3+1$ supersymmetric Bianchi IX model, the solution to the field equations have been shown to be mapped to the framework of a spinning billiard ball, and the algebra of the supersymmetry constriants and that of the Hamiltonian constranits has been related.
\\
All the quantum analysis in the above is based on the eigenvalue equation arising from the geometrical features of the UPHP. It is worth remarking, nonetheless, that it is reasonable to hypothesize that, at the quantum level, the features of the spacetime become discrete. Among all the possible models, where a discretized nature of spacetime is somehow hypothesized, a key-role is played by the so called loop quantum gravity models. Within this model, a discretized nature of space is implemented, and the resulting model in the minisuperspace representation, the so-called loop quantum cosmology, i.e. the asymptotic limit of loop quantum gravity models close to the cosmological singularity, accounts for the hypothesis according to which, in such a limit, the discretized nature of spacetime is appreciated when distances (which correspond to the early age of the universe) of the order of the Planck scale are reached. More precisely, the specific features of loop quantum gravity, and, in particular, of its 
application to cosmology, are such that the discretized nature of spacetime arises exactly when the Planck scale is taken into account.\\
The Hamiltonian framework within which the BKL paradigm can be implemented in Loop Quantum Gravity has been discussed in \cite{Ashtekar:2008jb} and \cite{Ashtekar:2011ck},, where a reduced phase space has been considered also within the specifications of this modified model.\\
The properties acquired by the Bianchi IX cosmological model in the asymptotic limit towards the cosmological singularity within the framework of loop quantum cosmology are presented in \cite{bojo2004}. In this analysis, the modifications arising from the discretized nature of spacetime are investigated. In particular, the discretized nature of spacetime close to the cosmological singularity is accounted for by the presence of discretized volume 'cells', whose facets have the size of the Planck scale. Furthermore, also the potential walls are described according to the same discretized nature. As a result, two different regimes are individuated. On the one hand, when the size of the universe is far greater then the Planck scale, no important modifications are induced on the chaotic behavior of the Bianchi IX model. On the contrary, when the size of the universe become comparable with the Planck scale, the discretized nature of spacetime here hypothesized strongly modifies the chaotic behavior proper 
of the Mixmaster universe. More in detail, the discretized nature of space allows for a change in the oscillating behavior of the scale factors because such an oscillation would become comparable with the Planck scale. This way, the never-ending sequence of Kasner oscillations obtained in the classical regime is no more a characterizing feature of the model, since the 'loop'-modified potential is shown to become progressively modified. Eventually, only a finite number of Kasner reflections on the billiard walls are possible.\\
A discussion of the model-independent features characterizing the properties of the universe at ages which are immediately following the Planck phase, and for which some of the ingredients which would allow for a reconstruction of the thermal history of the universe are also considered, is presented in
\cite{deOliveira:1999bj}, where the presence of chaos in the dynamics is analyzed.\\
\\
The quantum version of the model brings the problem of the interpretation of the wave function \cite{gibb} in quantum cosmology. In fact, the BKL paradigm, consisting in neglecting the space gradients, allows one to recover an arena with a finite number of degrees of freedom, for which the quantum implementation corresponds to a quantum-mechanical description rather than a field-theoretical description. In this respect, in \cite{Graham:1990jd}, the most relevant puzzling features of the quantum-mechanical description for the wave function of a generic universe (i.e. for a universe described by the most general solution to the Einstein field equations) are analyzed, and the absence of an 'observer' is pointed out.\\
On the UPHP, the reduced phase space construction is based on the assumption (deriving form the BKL conjecture itself) of elastic  bounces on the sides of the billiard table, and absence of friction on the billiard table, for which energy is conserved. From Hamiltonian theory of systems, this leads, at a classical level, to a standard definition of time. From the quantum point of view, however, the definition of time is still a controversial issue.\\
In this respect, in the Misner-Chitr\'e framework, the isotropic growth of the universe can be considered as a Hamiltonian time variable.\\
Within the framework of Loop quantum cosmology, on the other hand, the presence of a scalar field \cite{asht} \cite{asht1} \cite{isham1} can be interpreted, for the case of an FRW model, as an emergent relational time, as the scalar field is a monotonic function of the Hamiltonian time variable. This way, the presence of an 'observer', which defines all the 'thought experiments' in the foundations of quantum mechanics, is formally recovered by the interpretation of the scalar field as a relational time.\\
\\
More in general, the problem of time in quantum gravity is a long-standing issue, for which different solutions have been proposed. The square root of a differential operator, analyzed in the above, descends from an ADM decomposition of the dynamics \cite{adm}. This way, the definition of time in the quantum version of the foliation of the spacetime has been viewed as a tantalizing dilemma. A very important work is this direction has been performed by \cite{isham}, where the definition of time has been demonstrated to be possible both before the quantization and after the quantization.\\
%%%%%%%%%%%%%%%%%%%%%%%%%%%%%%%%%%%%%%%%%%%
\section{Domains\label{section2a}}
To better characterize the features of these billiards, their properties on the UPHP will be investigated.\\
The main features of some groups (whose properties will be used in the following) are here briefly revised. For the sake of a compact description, only the properties which are relevant in the analysis of the cosmological singularity will be treated, while a systematic classification can be found in \cite{terras}.\\
\\
%%%%%%%%%%%%%%%%%%%%%%%%%%%%%%%%%%
\subsection{The Modular group}
It consists of the two transformations
\begin{subequations}\label{sl}
\begin{align}
&T(z)=z+1,\\
&S(z)=-\tfrac{1}{z}
\end{align}
\end{subequations}
Its fundamental domain is one delimited by the sides $A_1$, $A_2$, $A_3$, defined by
\begin{subequations}\label{sl2}
\begin{align}
&A_1: \ \ u=-\tfrac{1}{2},\\
&A_2: \ \ u=\tfrac{1}{2},\\
&A_3: \ \ u^2+v^2=1.
\end{align}
\end{subequations}
The sides are identified as
\begin{subequations}
\begin{align}
&T: \ \ A_1\rightarrow A_2,\\
&S: \ \ A_3\rightarrow A_3,
\end{align}
\end{subequations}
such that
\begin{subequations}
\begin{align}
&Tz=z+1,\\
&Sz=-\frac{1}{z}.
\end{align}
\end{subequations}
%%%%%%%%%%%%%%%%%%%%%%%%%%%%%%%%%%%%%%%%%%%%%%%%%%%%%%%%%%%%%%%%
\subsection{The PGL(2,$\mathbb{Z}$) group}
Its fundamental domain consists of half that of Eq. (\ref{sl2}), and coincides with the small-billiard table (sketched in Figure \ref{figura2}, black thick solid line). It is generated by the transformations against the sides
\begin{subequations}\label{smalltrasf}
\begin{align}
&R_1(z)=-\bar{z},\\
&R_2(z)=-\bar{z}+1,\\
&R_3(z)=\tfrac{1}{\bar{z}}
\end{align}
\end{subequations},
which give the complete $z$ version of the small-billiard map $t$. The reflection laws in the $u^-u^+$ plane for the small billiard table as a function of the variable $u$ are therefore given by Eq. (\ref{smalltrasf}) evaluated on the $v=0$ axes, where the oriented endpoints of the geodesics are defined.
Because of the domain of this group is obtained by a desymmetrization of that of the symmetric domain (\ref{sl2}), it is not possible to identify sides in any manner.\\
It is straightforward to verify that the transformations (\ref{sl}) are obtained by a suitable composition of Eq. (\ref{smalltrasf}), i.e.
\begin{subequations}\label{comp}
\begin{align}
&T=R_2R_1,\ \ \ \ T^{-1}=R_1R_2,\\
&S=R_1R_3,\ \ S^{-1}=S,
\end{align}
\end{subequations}
while the converse is not true, it is not possible to obtain Eq. (\ref{smalltrasf}) from Eq. (\ref{sl}). In particular, Eq. (\ref{sl}) do not include complex conjugation, and can be interpreted as a composition of two (an even number of) reflections.\\
%%%%%%%%%%%%%%%%%%%%%%%%%%%%%%%%%%%%%%%%%%%%%%%%%%%%
\section{\label{section3}Billiard maps on the UPHP}
Given an epoch parametrized by the geodesics
\begin{equation}\label{geod}
\left(u-u_0\right)^2+v^2=r^2, \ \ \ \ u_0\equiv\frac{u^++u^-}{2}, \ \ \ \ r\equiv\frac{u^+-u^-}{2},
\end{equation}
transformations can be defined for the billiard trajectory parametrized by Eq. (\ref{geod}). The $xy$ kinds of epochs and eras are still defined by the values of $u^-$ and $u^+$, as explained in Section \ref{section2}, defined in Table II and Table III of \cite{Damour:2010sz}, and sketched in Figure \ref{figura3}.\\
It is more convenient, nevertheless, to examine the maps on the points $z=u+iv$ (which belong to a given geodesics (\ref{geod})) that imply the transformation laws for the geodesics (\ref{geod}), rather than the transformation laws for the geodesics themselves. In fact, the reduction of these maps on the $u$ axis will express the transformation law for the variables $u^+$ and $u^-$ in Eq. (\ref{geod}).\\ 
In this respect, it is useful to recall that, in \cite{Damour:2010sz}, the choice to coordinatize the reduced phase space by $u^-$ and $u^+$, rather than by $u_0$ and $r$, is due to the possibility, for the $(u^-,u^+)$ choice, to obtain subregions delimited by straight lines ('boxes') for the definition of $B_{xy}$ and $F_{xy}$.\\   
The interest in this analysis is due to the fact that the complete maps for the $z$ variable in the UPHP automatically take into account the way the roles of the Kasner exponents are quotiented. As a result, the proper understanding of the symmetry quotienting of the Kasner exponents allows one to impose the correct identifications of the boundaries of the billiard tables. These identifications then define the (sub)group structure of the maps.\\
In particular, the maps that will be considered are the implementation of the maps $\mathcal{T}$ Eq. (\ref{curv}), $T_{\rm BKL}$ Eq. (\ref{BKL}), $\mathbf{T}$ Eq. (\ref{cblksks}) and $t$ Eq. (\ref{t}) for the complete $z$ variable in the UPHP.  
\subsection{The big-billiard map}
The big billiard is defined on the domain defined in Eq. (\ref{a}) and sketched in Figure \ref{figura1}. No identification of the sides is required, as the big billiard map $\mathcal{T}$ allows one to follow the dynamics throughout the corners, and to keep track of the orientation.\\
The big-billiard map $\mathcal{T}$ for the variable $z=u+iv$ can be obtained from Eq. (\ref{curv}), and expressed as a suitable composition of the elements of Eq. (\ref{smalltrasf}), i.e. 
\begin{subequations}\label{bbz}
\begin{align}
&A: \ \ z\rightarrow -\bar{z}, \ \ \ \ A\equiv R_1,\\
&B: \ \ z\rightarrow -\bar{z}-2, \ \ \ \ B\equiv R_1(R_2R_1)(R_2R_1)\\
&C: \ \ z\rightarrow -\tfrac{\bar{z}}{2\bar{z}+1}, \ \ \ \ C\equiv(R_1R_3)(R_2R_1)(R_2R_1)R_3, 
\end{align}
\end{subequations}
where, for the sake of compact notation, the original names given in Eq. (\ref{curv}) have been kept; anyhow, no confusion should arise, as these transformations will act here and in the following on the $z$ variable.\\ 
As a result, one sees that the transformations (\ref{bbz}) consist of an \textit{odd} number of reflections, under which the dynamics has to be invariant.\\
\\   
\subsection{The Kasner-quotiented big billiard maps}
The $S_3$-quotienting of the big-billiard is obtained by defining the BKL maps and the CB-LKSKS map on the $(u,v)$ plane for the appropriate $z-{\rm BKL}$ map. The choice of identification of boundaries of the corresponding domain will play a crucial role in the analysis of the maps.\\
\\
To define the symmetry-quotienting maps, one first needs the $z$ version of the Kasner transformations, say $K_i$. The Kasner transformations $K_i$ for the variable $z=u+iv$ are given in Table \ref{table1}, where they are also expressed as a suitable composition of the elements of Eq. (\ref{smalltrasf}). For $v=0$, the $k_i$ Kasner transformations for the $u$ variable alone of Table I and Table VI in \cite{Damour:2010sz} are recovered. The transformations $k_i$ also act diagonally on the variables $(u^+, u^-)$ of the reduced phase space. The determinant of the $k_i$'s is equal to $1$ for the epoch types oriented in the clock-wise sense (i.e. $ba$, $ac$ and $cb$, 'parallel' to $ba$), and to $-1$ for the epoch types oriented in the counter-clock-wise sense (i.e. $ab$, $bc$ and $cb$, 'antiparallel' to $ba$). Similarly, the $K_i$'s for the epoch types 'parallel' to $ba$ contain an even number of reflections (and do not imply complex conjugation), while those for the epoch types 'antiparallel' to $ba$ contain an 
odd number of reflections. This way, as given in Table \ref{table1}, the Kasner transformations $K_0$, $K_2$ and $K_4$ are \textit{rotations}, while the Kasner transformations $K_1$, $K_3$ and $K_5$ are \textit{reflections}.\\
%%%%%%%%%%%%%%%%%%%%%%%%%%%%%%%%%%%%%%%%%%%%
\begin{table}
\begin{center}
    \begin{tabular}{ || l || l || l | l | }
    \hline
      $ca\rightarrow ba$ & $K_1$ & $z\rightarrow 1/\bar{z}$ & $ R_3$\\ \hline
      $ac\rightarrow ba$ & $K_2$ & $z\rightarrow -(1+z)/z$ & $(R_1R_2)(R_1R_3)$\\ \hline
      $bc\rightarrow ba$ & $K_3$ & $z\rightarrow -\bar{z}/(\bar{z}+1)$ & $(R_1R_3)(R_2R_1)R_3$\\ \hline
      $cb\rightarrow ba$ & $K_4$ & $z\rightarrow -1/(z+1)$ & $(R_3R_1)(R_2R_1)$ \\ \hline
      $ab\rightarrow ba$ & $K_5$ & $z\rightarrow -\bar{z}-1$ & $R_1R_2R_1$\\ \hline
      $ba\rightarrow ba$ & $K_0$ & $z\rightarrow z$ & $I$ \\ \hline
    \end{tabular}
\end{center}
\caption{\label{table1} The Kasner transformations.\newline
\textbf{Right Panel}. The action of the Kasner transformations. They act on the set of the Kasner exponents $a$, $b$, $c$ by suitable permuting them. For the preferred order $a,b,c$, the Kasner transformations have the effect of mapping $xy$ epochs in $ba$ epochs.\newline
\textbf{Left Panel}. The Kasner tranformations $K_i$ acting on the variable $z=u+iv$ of the UPHP. They can be expressed as a suitable composition of the reflections (\ref{smalltrasf}). Evaluated at $v=0$, the Kasner transformations $K_i$'s reduce to the $k_i$'s of \cite{Damour:2010sz} for the variable $u$ and for the reduced phase space variables $u^{\pm}$.}
\end{table}

The BKL map and the CB-LKSKS map for the variable $z_{\rm BKL}$ is obtained by a suitable composition of the big-billiard map (\ref{bbz}) and the Kasner transformations $K_i$ given Table \ref{table1} for the (preferred $ba$-type epochs).\\
\\
%%%%%%%%%%%%%%%%%%%%%%%%%%%%%%%%%%%%%%%%%%%%%%%%%%%
\subsection{The BKL epoch map on the UPHP}The definition of the BKL epoch map can be constructed as follows, for the variable $z_{\rm BKL}$. For the case of a $3$-epoch $ba$ era (which consists of the epochs $z^1_{ba}$, $z^2_{ab}$ and $z^3_{ba}$) , the following compositions hold:
\begin{subequations}\label{BKLzetaepoch}
\begin{align}
&z^1_{ba}, \ \ \ \ \ \ \ \ \ \ \ \ \ \ \ \ \ \ \ \ \quad\quad\quad\quad z^{1}_{\rm BKL}\equiv K_0z^1_{ba}\\
&z^2_{ab}=A(z^1_{ba})=-z^1_{ba},\ \ \ \ \ \ \ \ \ \ \ z^2_{\rm BKL}=K_5A(z^1_{ba})=z^1_{ba}-1\equiv z^{1}_{\rm BKL}-1\\
&z^3_{ba}=BA(z^1_{ba})=z^1_{ba}-2,\ \ \ \ \ \ z^3_{\rm BKL}=K_0(z^3_{ba})=z^1_{ba}-2\equiv z^{1}_{\rm BKL}-2
%&z^1_{ac}=ABA(z^1_{ba})=-z^1_{ba}+2,\ \ z^{1'}_{\rm BKL}=K_2ABAz^1_{ba}.
\end{align}
\end{subequations}
This way, the BKL epoch map is generated by the (composition of) translations $T^{-m}$ for suitable values of $m$, $0\le m\le n-1$. 
For the variable $z_{\rm BKL}$, the BKL epoch map then reads
\begin{equation}\label{bklzepoch}
z_{\rm BKL}\rightarrow T^{-1}z_{\rm BKL}-1\rightarrow T^{-2}z_{\rm BKL}\rightarrow...
\end{equation}
and, given $z_{\rm BKL}=u_{\rm BKL}+iv_{\rm BKL}$, one easily obtains
\begin{equation}\label{bkluvepoch}
u\rightarrow u-1,\ \ \ \ v\rightarrow v.
\end{equation}
\\
As already specified, the BKL epoch map identifies the the $a$ and $b$ Kasner coefficients and keeps that of the $c$ coefficient as fixed. According to the roles of the billiard walls in defining the oscillatory behavior of the $a$ and $b$ coefficients by the corresponding reflections $A$ and $B$ in Eq. (\ref{bbz}) while the $c$ side of the billiard has not been reached yet, it is natural to require boundary identifications for the big-billiard table as
\begin{subequations}\label{emptbound}
\begin{align}
&b\rightarrow a, \ \ z\rightarrow z+1,\\
&c\rightarrow c, \ \ z\rightarrow \tfrac{z}{2z+1}-1.
\end{align}
\end{subequations}

As the BKL map consists in describing the dynamics by mapping all the epochs in $ba$ epochs, it is fully consistent to remark that Eq. (\ref{BKLzetaepoch}) consists of an even number of reflections, that compose as translations. The 'bouncing' behavior typical of billiards is then completely eliminated.\\
\\
It is interesting to remark that the BKL epoch map for the $z$ variable on the UPHP is defined by the boundary identifications (\ref{emptbound}), which consistently describe the symmetry-quotienting of the roles of the $a$ and $b$ Kasner exponents, while the $c$ Kasner exponent is left unchanged, during a given BKL epoch. In the reduced phase space $(u^-,u^+)$, contrastingly, the BKL epoch map for the $u^{\pm}$ variable is defined by the behavior $u\rightarrow u-1\rightarrow...$ only, for which it is in principle difficult to trace back the symmetry-quotienting mechanism.\\
Furthermore, the domain with the boundary identifications (\ref{emptbound}) takes into account both the quotienting of the roles of $a$ and $b$ and the identification of $c$ with itself during the same BKL epoch, in the UPHP for the variable $z$. On the other hand, the BKL epoch map in the $(u^-,u^+)$ reduced phase space takes into account only for the symmetry quotienting of the roles of $a$ and $b$ (while the side $c$ of the billiard has not been reached yet), because this description involves only the BKL $u$ variable and not the complete dynamics.\\
\\   
%%%%%%%%%%%%%%%%%%%%%%%%%%%%%%%%%%%%%%%%%%%%%%%%%
\subsection{The BKL era-transition map and the CB-LKSKS map on the UPHP}The next step is to define the BKL era-transition map in the UPHP, and then to express the CB-LKSKS as a sequence of BKL era-transition maps for the variable $z$.\\
As in the definition of the BKL epoch map, for the case of a $3$-epoch $ba$ era (which consists of the epochs $z^1_{ba}$, $z^2_{ab}$ and $z^3_{ba}$) followed by an $ac$ epoch $z^1_{ac}$, the succession (\ref{BKLzetaepoch}) is continued by
\begin{equation}\label{BKLzetaera}
z^1_{ac}=ABA(z^1_{ba})=-z^1_{ba}+2,\ \ z^{1'}_{\rm BKL}=K_2ABAz^1_{ba}.
\end{equation}
(The case of a $ba$ era containing an even number of epochs can be treated analogously and admits the very same composition). Eq. (\ref{BKLzetaera}) describes the transition to the new era $z'$, where the role of $c$ exponent is taken by either $a$ or $b$ (whose role has already been quotiented). It is straightforward to remark that the BKL era-transition map (\ref{BKLzetaera}) contains an \textit{odd} number of reflections, i.e. it implies complex conjugation. For the sake of compactness, this features will be discussed after the definition of the CB-LKSKS map.\\
\\ 
As a result, the complete BKL (epoch and era-transition) map for the $z_{\rm BKL}$ variable reads
\begin{equation}\label{BKLz}
z\rightarrow T^{-1}z\rightarrow ... \rightarrow T^{-n+1}z\rightarrow z'\equiv\frac{1}{\bar{z}-n+1}-1\equiv T^{-1}SR_1T^{-n+1}z
\end{equation}
(For the sake of easy notation, the symbols $T$ and $S$ are used in Eq. (\ref{BKLz}) as a 'shorthand' for their expression (\ref{comp}) as functions of Eq. (\ref{smalltrasf}). Furthermore, the subscript ${}_{\rm BKL}$ has been omitted to keep the notation compact; it will be restored when needed to avoid confusion).\\
Given $z_{\rm BKL}=u_{\rm BKL}+iv_{\rm BKL}$, one easily obtains the era-transition map
\begin{equation}\label{BKLuv}
u\rightarrow -\frac{(u-n)(u-n+1+v^2)}{(u-n+1)^2+v^2},\ \ \ \ v\rightarrow \frac{v}{(u-n+1)^2+v^2}.
\end{equation}
\\
Neglecting the succession of epochs in each era, the following expression for the CB-LKSKS map $\mathbf{T}$ is found
\begin{equation}\label{cblksksz}
\mathbf{T}z=\frac{1}{\bar{z}-n+1}-1\equiv T^{-1}SR_1T^{-n+1}z,
\end{equation}
which is interpreted as a succession of BKL era-transition maps for the variable $z$. As analyzed in the above, it is then straightforward to recognize that the CB-LKSKS map for the variable $z$ will contain only an \textit{odd} number of reflections.\\
\\
Although the BKL map and the CB-LKSKS do not take into account the bounces on the billiard walls (that characterize the unquotiented dynamics), and is, on the contrary, based on a description involving always the same ($ba$) kind of epochs and eras, it is not possible to express the BKL map as containing only an even number of reflections at each step.\\
In other words, the BKL transition map and the CB-LKSKS map define the new radius $r'$ and the new $u$ value for the center $u_0'$, $v=0$ by means of reflections, although these maps always generate epochs and eras of the $ba$-type.\\
On the one hand, one sees that the part of the map containing a reflection is the era-transition map, where the Kasner exponent $c$ has to be taken into account. The fact that the role of the $c$ Kasner exponent in the BKL map is different form those of $a$ and $b$ could look unexpected, if one considers that the bounces on the three sides $a$, $b$, and $c$ of the big billiard are described by the laws (\ref{bbz}), which contain all an odd number of reflections- as all bounces are just defined as reflections against the billiard walls.\\
On the other hand, the reasons for this behavior have to be looked for in the difference between the boundary identification (\ref{emptbound}) and the way the BKL era-transition map (\ref{BKLzetaera}) quotients the roles of the Kasner exponents.\\
In fact, as already analyzed, the BKL epoch map identifies the exponents $a$ and $b$, while the $c$ Kasner exponent undergoes a monotonic evolution.\\ On the contrary, the BKL era-transition map and the CB-LKSKS identify the $c$ Kasner exponent with either $a$ or $b$ (already quotiented in the epoch map). As a result, these maps have to be defined on a domain with the following boundary identification
\begin{subequations}
\begin{align}
&b\rightarrow a,\ \ \ \ z\rightarrow z+1\label{refl}\\
&c\rightarrow a,\ \ \ \ z\rightarrow -1+\frac{1}{\bar{z}-1}\label{invers}
\end{align}
\end{subequations} 
which are inequivalent to the boundary identification (\ref{emptbound}) for the BKL epoch map. From a geometrical point of view, the transformation (\ref{invers}) is the inversion of the side $c$ with respect to the circle $(u-1)^2+v^2=1$ (which acts a a symmetry line of the big billiard, on which subdominant symmetry walls are defined).\\
Given all this, it can be useful to note that, although the BKL era-transition map and the CB-LKSKS map contain by construction on odd number of reflections, the intermediate step of the BKL epoch map is a necessary ingredient in the definition of the quotiented era maps. Furthermore, it is then possible to identify in the complete BKL map the two different parts, i.e. the epoch map containing no reflections, and the era-transition map containing reflections.\\      
\\
A remark is now in order. As explained in the above, any $xy$ epoch (or era) can be mapped as one of the $ba$ type by the suitable Kasner transformation $K_i$, and the corresponding $z_{\rm BKL}$ will be defined as $z_{\rm BKL}\equiv K_iz_{xy}$. Although the suitable $K_i$ may contain an odd or an even number of reflections (as illustrated in the above, and in Table \ref{table1}), the considerations developped until here hold for the variable $z_{\rm BKL}$ itself, no matter if in its derivation $z_{\rm BKL}\equiv K_iz_{xy}$ an odd number of reflections can be present through $K_i$. In fact, in that case, the only relevant physical information would be that the $xy$ type of epoch is 'antiparallel' to the $ba$ type, without affecting the behavior of the map for the BKL variable $z_{\rm BKL}$. In fact, the choice of the $ba$ type of epochs as a preferred one is merely do to the corresponding 'treatable' expressions (\ref{BKLz}) and (\ref{cblksks}), rather than to any intrinsic physical meaning.
%%%%%%%%%%%%%%%%%%%%%%%%%%%%%%%%%%%%%%%%%%%%%%%%%%%%%%%%
\subsection{The small billiard BKL map in the UPHP\label{smallbkl}}
As remarked above, and as explained in \cite{Damour:2010sz}, there is a mismatch in the definition of a BKL or CK-LKSKS map for the small billiard in the reduced phase space.\\
Anyhow, as remarked in \cite{Damour:2010sz}, it is possible to reconstruct the complete map from the reduced phase space by considering all the Kasner transformations, given in Table \ref{table1}, and then by considering the following.\\

It is important to recall that, in the $u^+u^-$ description of the dynamics, trajectories (i.e. epochs) are identified with the oriented endpoints (on the $v=0$ line) of the (extended) geodesics, where these trajectories lie.\\
The dynamics of each era is described by the behavior of the scale factors. In particular, for a generic era $xy$ in the $B_{xy}$ region of the epoch hopscotch court, the two scale factors $x$ and $y$ oscillate between zero and a maximum value at each epoch, while the $z$ scale factor (corresponding to the side of the billiard which is not hit during the era) has a monotonic evolution. During the $xy$ oscillation, the zero of a scale factor corresponds to the moment when the billiard ball collides on a gravitational wall, say $x$, such that the other scale factor, say $y$ reaches its maximum value, and vice versa. During each epoch, by construction, the two oscillating scale factors $x$ and $y$ cross once, when their values are equal. In the meanwhile, the non-oscillating scale factor $z$ undergoes a monotonic evolution, and may (but not necessarily has to) cross the other two. The beginning of an era is in fact defined as the moment when the non-oscillating scale factor $z$ changes its slope, but it can 
cross the other two scale factors in several ways, as it is straightforward to verify.\\

As a result, the set of epochs belonging to each eras can be further classified according to these properties.\\
The symmetry walls, that enclose the small billiard, can be interpreted as `symmetry lines' for the big billiard whose meaning is suggested by their geometrical construction: each of these three lines represents the points of the billiard table, where one of the scale factors equals another. As a consequence, the crossing of a symmetry line by an epoch singles out the point of that epoch, where the two scale factors are equal. It is easily verified that an epoch can cross one, two or three symmetry lines. This feature is illustrated in Figure \ref{figura1new}.\\

\begin{figure*}[htbp]
\begin{center}
\includegraphics[width=0.7\textwidth]{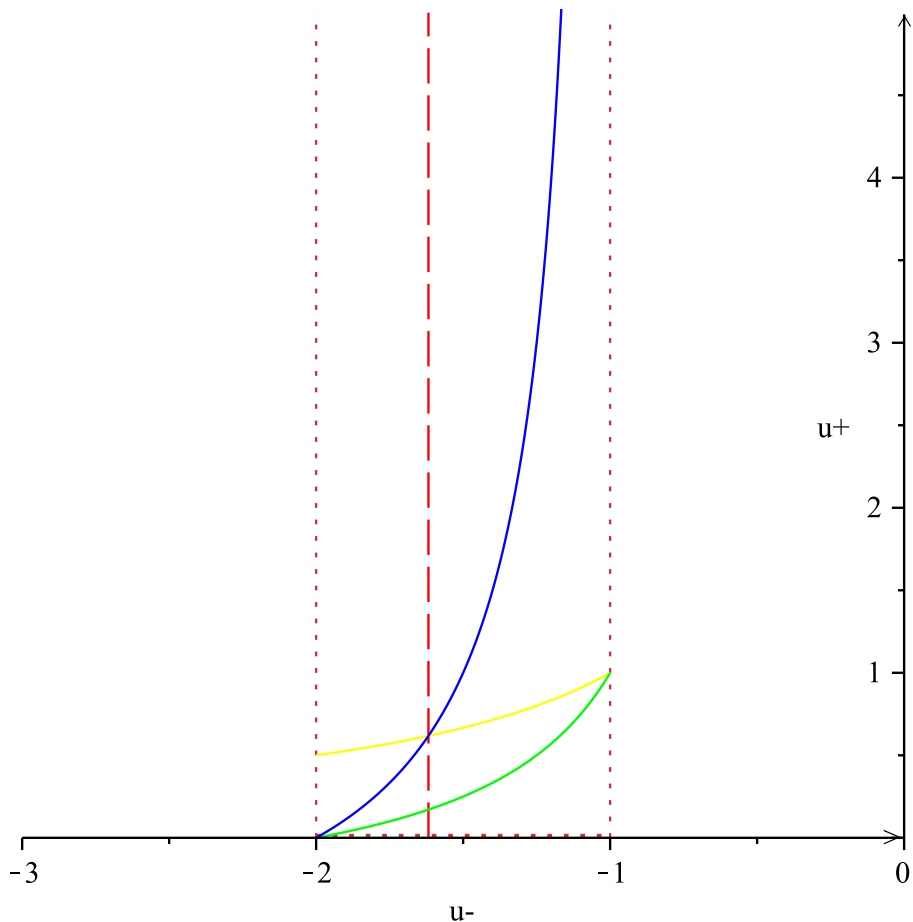}
\caption{\label{illustraz3} The subregions of the phase space of the $F_ba$ box of the billiard hopsotch court in the $(u^-, u^+)$ plane, where the BKL map acquires different features with respect to its content of Weyl reflections, and the function $u_\alpha$ [green (gray) line], $u_\beta$ [yellow (light gray) line] and $u_\gamma$ [ blue (dark gray) line] are plotted. The $F_{ba}$ box is delimited by the orange dotted line. The value $u^-=-\Phi$ divides the $F_{ba}$ starting box, as explained in Table \ref{table9} and in the text.}
\end{center}
\end{figure*}

Each $B_{xy}$ region of the epoch hopscotch court can be therefore further divided into subregions $S^m_{xy}$, according to the behavior of the scale factors, which corresponds to the ordered crossings of the symmetry lines. This regions are given in Table \ref{table8}, where the different kinds of 'crossings' are explicitly described according to the order they happen, and referred to by the index ${}^m$.\\
%%%%%%%%%%%%%%%%%%%%%%%%%%%%%%%%%%%%%%%%%%%%%%%%%
\begin{table}
\begin{center}
    \begin{tabular}{ || l || l | }
    \hline
    $ S^1_{xy} $ & $\ \ \ \ \ \  x=y $\\ \hline
    $S^2_{xy}$ & $\ \ y=z,\ \ \ \ x=y$\\ \hline
    $S^{2'}_{xy}$ & $\ \ x=y, \ \ \ \ x=z$\\ \hline
    $S^{3}_{xy}$ & $y=z,\ \ x=y,\ \ x=z$\\ \hline
    $S^{3'}_{xy}$ & $x=z,\ \ x=y,\ \ y=z$\\ \hline
    \end{tabular}
\end{center}
\caption{\label{table8} Dynamical subregions of the epoch hopscotch court.}
\end{table}\\
%%%%%%%%%%%%%%%%%%%%%%%%%%%%%%%%%%
It is crucial to remark that the five $S^m$ regions do not coincide with the partition of $B_{xy}$ into $F_{xy}$ and $L_{xy}$.\\
Let's specify this analysis to one particular region of the hopscotch court, $B_{ba}$, for which the details are listed in Table \ref{table9}, the functions $u_\alpha$ and $u_\beta$ are defined in Eq. (\ref{table4}), and $u_\gamma:\ \ u^+=-\tfrac{u^-+2}{u^-+1}$; the 'Golden Ratio' is approximated as $\Phi\sim 1,618$. The two functions $u_\alpha$ and $u_\beta$ cross at the point $u^-=-\Phi$ and $u^+=\phi$, where the 'Small Golden Ratio' $\phi$ is defined as $\phi+1=\Phi$. The corresponding subregions for the other kinds of eras are obtained from those given in Table \ref{table9} by applying the corresponding inverse Kasner transformation of Table \ref{table1}, i.e. $S^m_{ca}=K^{-1}_1S^m_{ba}$.\\
%%%%%%%%%%%%%%%%%%%%%%%%%%%%%%%%%%%%%%%%%%%%%%%%%%%%%%%%%%%%%%%%%
\begin{table}
\begin{center}
\begin{tabular}{ || l || l | }

  \hline
  
   $S^1_{ba} $  & $u^-<-\Phi,\ \ u^+>u_\alpha$   \\ 
      &  $-\Phi<u^-<-1,\ \ u^+>u_\gamma$ \\
  \hline
  $S^2_{ba} $  &  $-\Phi<u^-<-1,\ \ u_\alpha<u^+<u_\gamma$ \\
  \hline
  $S^{2'}_{ba} $  & $u^-<-2,\ \ 0<u^+<u_\alpha$   \\ 
      &  $-2<u^-<-\Phi,\ \ u_\gamma<u^+<u_\alpha$ \\
  \hline
  $S^{3}_{ba} $  & $-2<u^-<-\Phi,\ \ u_\gamma<u^+<u_\beta$   \\ 
      &  $-\Phi<u^-<-1,\ \ u_\alpha<u^+<u_\beta$ \\
  \hline
  $S^{3'}_{ba} $  & $-2<u^-<-1,\ \ 0<u^+<u_\beta$  \\ 
  \hline

\end{tabular}
    %\label{tab:secondlevel}
\end{center}
\caption{\label{table9} The crossing subregions $S^m_{ba}$ for the $B_{ba}$ portion of the epoch hopscotch court. This Table specifies the details of Figure \ref{illustraz3}}
\end{table}
%%%%%%%%%%%%%%%%%%%%%%%%%%%%%
The comparison of the dynamics of the small billiard with that of the big billiard is consistent only in those regions of the $u^+,u^-$ plane, where it is possible to identify trajectories. In fact, for the big billiard, the $u^+u^-$ plane describes collisions on gravitational walls. As a consequence, it is consistent to consider only the regions of the $u^+u^-$ plane of the small billiard, where collisions on gravitational walls are described, i.e. the $BG$ and the $RG$ regions. It is straightforward to verify that the $BG$ and the $RG$ regions of the small billiard restricted phase space exactly correspond to the regions $S^1_{ba}$, $S^2_{ba}$, $S^{2'}_{ca}$, $S^3_{ca}$, $S^{3'}_{ca}$ depicted in Figure \ref{illustraz3} and specified in Table \ref{table9}. By considering the proper Kasner transformations, it is therefore possible to recover exactly one $B_{xy}$ region of the epoch hopscotch court, i.e. $u\rightarrow1/u$ for $-1<u^-<0$, which allows one to map $S^{2'}_{ca}$, $S^3_{ca}$, $S^{3'}_{ca}$ into $S^{
2'}_{ba}$, $S^3_{ba}$, $S^{3'}_{ba}$.\\
\\
%%%%%%%%%%%%%%%%%%%%%%%%%%%%%%
%%%%%%%%%%%%%%%%%%%%%%%%%%%%%%%
\\
Then, the CB-LKSKS map for the small billiard, $t_{CB-LKSKS}$, is defined by
\begin{subequations}\label{tsb}
\begin{align}
&t^{1,2}z=T^{-1}SR_1T^{-n+1}z,\ \ {\rm for} (u^+, u^-)\in S^{1}_{ba} {\rm and} (u^+, u^-)\in S^{2}_{ba},\label{tsb1}\\
&t^{2',3,3'}z=T^{-1}SR_1T^{-n+1}R_3z,\ \ {\rm for} (u^+, u^-)\in S^{2'}_{ba}, (u^+, u^-)\in S^{3}_{ba}, {\rm and} (u^+, u^-)\in S^{3'}_{ba}\label{tsb2}.
\end{align}
\end{subequations}
The transformation $R_3$ in the second $t_{CB-LKSKS}$ map is given by the $K_1$ Kasner transformation needed to reconstruct the $B_{ba}$ region in the restricted phase space.\\
It is immediate to verify that the two maps contain a different number of reflections.\\
\\
As a result, the meaning of the $t_{CB-LKSKS}$ map can be inferred from its action on the description of the dynamics.\\
The transformation  $R_3$ in Eq. (\ref{tsb2}) quotients out the order of red crossings in the succession of eras. Small billiard eras are defined in \cite{Damour:2010sz} by the hit on the red wall, and a billiard map for this phenomenon has been pointed out to be much more 'intractable' than a CB-LKSKS map because the regions of the small billiard phase space are not defined by straight lines, and there exists no transformation able to map the curved domains into square boxes. This is consistent with the fact that the small billiard domain is not symmetric, in the sense that there exist no transformation able to identify its sides, as already pointed out.\\
The maps (\ref{tsb}) avoids the problem of identifying sides of the small billiard by identifying the kind and the order of the 'crossings' of the symmetry lines, for which the definition of small billiard eras looks in principle lost. Anyhow, by using the map (\ref{tsb}), it is always possible to switch from the small billiard era description, where the hit on the red wall defines an era, to the quotiented big-billiard description, where BKL eras are described by the Poincar\'e return map on the (identified) green wall(s). In other words, by the definition of a Poincar\'e return map on the green walls, any information about the crossing of the symmetry lines of the billiard is disregarded. Anyhow, as shown in Table \ref{table9}, this information can be recovered also in the BKL quotiented billiard. Also in the BKL-quotiented billiard, the subregions $S^m_{sm}$ are delimited by curved lines and are not box-like domains, and there exists no transformation able to map these regions into boxes.\\
From this point of view, one can infer also that, for the sake of a Poincar\'e return map on the green walls, the kind and the order of crossings is unimportant. In fact, the $\tau$ time evolution of the scale factors defines the change of eras as the point where the non-oscillating scale factor inverts its slope. From the BKL map point of view, this point defines when two different sides have to be identified, while, from the CB-LKSKS map point of view, all sides are always identified, as described in Table \ref{table5}.\\
%%%%%%%%%%%%%%%%%%%%%%%%%%%%%%%%%%%%%%%%%%%%%%%%
\begin{table}
\begin{center}
    \begin{tabular}{ | l | l | l | }
    \hline
    $ {\rm big\ \ billiard} $ & $ {\rm epoch\ \ map} $ & $ {\rm era\ \ map} $\\ \hline
    $ a\rightarrow a $ &      $ a\rightarrow a $ & $a\rightarrow a$\\
    $ b\rightarrow b $ &      $ b\rightarrow a$ & $b\rightarrow a$\\  
    $ c\rightarrow c $ &       $  c\rightarrow c          $ & $c\rightarrow a$\\ \hline
    \end{tabular}
\end{center}
\caption{\label{table5} Comparison of the unquotiented big- billiard map, the BKL epoch map, and the CB-LKSKS map (consisting of the composition of successive BKL era-transition maps), according to their different actions in quotienting the roles of the Kasner exponents, which is described by the different identifications of the boundaries of the billiard table.}
\end{table}
%%%%%%%%%%%%%%%%%%%%%%%%%%%%%%%%%%%%%%%%%%%%%%%%%%
In the unquotiented dynamics, the change of era (i.e. the change of the corner where the oscillation takes place) is always accompanied with the crossing of a trajectory of a red symmetry line, and this crossing is not relevant for the point where the non-oscillating scale factor changes its slope, as one sees from Table \ref{table9}. For this reason, an exact correspondence between the length of big billiard eras and small billiard eras is in principle lost, but can be recovered case by case, if needed, by gathering all the information together. 
%%%%%%%%%%%%%%%%%%%%%%%%%%%%%%%%%%%%%%%%%%%%%%%%%%%%%%%%%%%%%
\subsection{More about the era-transition map on the UPHP}
By means of all these considerations, the presence of a (complex) reflection in the maps (\ref{BKLz}) and (\ref{cblksksz}) can be further understood as follows.\\
Following the example of a $3$-epoch era of the $ba$ type, followed by an $ac$ epoch, two different subcases are possible. In particular, either the $ac$ epoch is the first epoch of an $ac$ era, i.e. according to the succession
\begin{equation}\label{firstcase}
ba \rightarrow ab \rightarrow ba \rightarrow ac \rightarrow ca \rightarrow ...
\end{equation}
or there is a $1$-epoch era of the $ac$ type followed by a $cb$ epoch, i.e. according to the succession  
\begin{equation}\label{secondcase}
ba \rightarrow ab \rightarrow ba \rightarrow ac \rightarrow cb \rightarrow ...
\end{equation} 
By construction, one can easily verify that, in the first case, i.e, according to the succession (\ref{firstcase}), only one red symmetry line is crossed, i.e., in the small billiard version, the red wall is hit only once. Anyhow, the red symmetry line can be crossed either during the last $ba$ epoch, or during the first $ac$ epoch.\\
On the contrary, in the case of the succession (\ref{secondcase}), the red symmetry lines of the billiard can be crossed either one or three times.\\
These considerations explain that
\begin{itemize}
	\item there is a mismatch in the comparison of big-billiard eras and small-billiard eras, i.e. the red line can be crossed during different epochs;
	\item there is a mismatch also in the number of eras counted on the big billiard dynamics and on the small-billiard dynamics, i.e. the red symmetry line can be crossed even three times during the transition from different eras (recall that the crossing of a red symmetry line defines a small-billiard era);
	\item there is a (complex) reflection in the era-transition map on the UPHP, i.e. the number of crossings of the red symmetry lines corresponds to the number of complex transformations of the small-billiard dynamics;
	\item the role of the Kasner transformations in Table \ref{table1} can be interpreted as reflections with respect to the symmetry lines; the fact that half of them is complex and half of them is real is due to the further choice of re-orienting the type of eras with respect to the preferred $ba$ type. 
\end{itemize} 
%%%%%%%%%%%%%%%%%%%%%%%%%%%%%%%%%%%%%%%%%%%%%%%%%%%%%%%%%%
\subsection{Comparison of the maps on the UPHP}
According to all the steps followed in the above, a paradigm can be implemented, according to which it is possible to reconstruct all the maps starting from the knowledge of the desymmetrized fundamental domain, corresponding to the more general inhomogeneous case, i.e. the geometrical symmetry-quotienting.\\

The paradigm can be summarized as follows
\begin{itemize}
	\item take twice the desymmetrized domain, and obtain a symmetric domain with suitable boundary identification, corresponding to the fundamental domain of the group (\ref{sl});
	%\item build the $\theta$ congruence subgroup for the symmetric group (\ref{sl}), and suitably rescale it;
	\item the unquotiented big billiard map is obtained by interpreting its domain as a suitable congruence subgroup of the desymmetrized domain without any boundary identification;
	\item the BKL epoch map for the big billiard is defined for a domain where only two sides are identified (in the $ba$ case) by translation, and the other is identified with itself. The BKL epoch map for the big billiard is the generated by the iteration(s) of the identification;
	\item the BKL era-transition map for the big billiard is obtained by identifying all the sides of the billiard onto (a preferred) one, and is generated by (a suitable composition of) these identifications;
	\item the CB-LKSKS map for the big billiard is obtained by the iteration(s) of the BKL era-transition map;
	\item the CB-LKSKS map for the small billiard is obtained from that of the big billiard by inserting the reflection $R_3$ in the CB-LKSKS map where needed, and has the meaning of identifying the green sides of the big billiard by neglecting the information on the crossing of the symmetry lines; 
	\item the maps for the reduced phase space ($u^-u^+$) are obtained by evaluating the maps in the UPHP on the absolute $v=0$. This maps act diagonally on each variable $u^-$ and $u^+$.  
\end{itemize}
On the other hand, the understanding of the BKL map in the ($u^-,u^+$) plane allows one to understand the exact partition of the dynamics into epochs and eras by means of the 'boxes' $B_{xy}$ and $F_{xy}$ (given in Table II and Table III of \cite{Damour:2010sz}) and by means of the precise relations between the number of epochs in each era and the value of the variables $u^+_F$, as given in \cite{Damour:2010sz}, for which it is possible to work out the relation $n_{ba}-1=[u^+_{ba}]$ used in Eq. (\ref{BKL}).
The main features of these maps are summarized in Table \ref{table5}, where the different symmetry quotienting mechanisms of the Kasner exponents are illustrated according to the different identifications of the sides of the billiard table. The three maps are defined on the \textit{same} domain, given in Figure \ref{figura1}.\\
\\
%%%%%%%%%%%%%%%%%%%%%%%%%%%%%%%%%%%%%%%%%%%%%%%%%%%%%
\section{The big billiard domain and the small billiard domain in Early Cosmology\label{congruence}}
In \cite{Fleig:2011mu}, Eq. (32), a compact formula is found for the volume of the small billiard in any number of spacetime dimensions. The volume is found as a volume integral on the appropriate region: In the case of cosmological billiards, this is the Weyl chamber of the group which defines the dynamics of the small billiard. It is defined as the simplex in the (generalized) UPHP, where the Euclidean definition of simplex is modified according to the non-trivial metric (which defines a non-trivial volume element) of the UPHP. This definition of generalized simplex is shown to encode all the symmetries that characterize, in the last analysis, the solution to the Einstein field equations for the pertinent case.\\
\\
The shape of this fundamental domain is then defined, on the generalized higher-dimensional UPHP, as the portion of space enclosed by the generalized simplex and the unit hypersphere centered on the generalized absolute (generalized hyperplane) of the UPHP. The generalized simplex, as far as the description in terms of cosmological billiards is concerned, consists of a portion of a gravitational wall, and portions of the symmetry walls. The portion of the gravitational wall is one among those that enclose the cusp at $v=\infty$, while the portions of the symmetry walls are among those that intersect this cusp.\\
This construction directly generalized that of the small billiard in $4=3+1$ spacetime dimensions, where the $G$ wall and the $B$ one are defined. Here, the $G$ wall defines the cusp, while the $B$ wall bisects the cusp.\\
On these walls, affine reflections \cite{Kleinschmidt:2009cv} are defined.\\
\\
The portion of the hypersphere, on which the remaining parts of the integration boundaries are defined, is the direct generalization, on its turn, of the $R$ wall for the case of the small billiard. On these hyper surface, Weyl reflections \cite{Kleinschmidt:2009cv} are defined, for the portions of the symmetry walls which define the Weyl chamber but do not intersect the cusp.\\
\\
As a result, the volume of the domain of the big billiard table is proportional to that of the small one, multiplied by a prefactor, which equals the order of the congruence subgroup, which defines the big billiard with respect to the small billiard.\\
\\
From the analysis of Eq. (\ref{bbz}), one finds\footnote{Among the several definitions adopted in the literature, the definition of \cite{elliptic} has been followed.} that this congruence subgroup is $\Gamma(2)$, whose index is $6$. This is equivalent to the construction for which the big billiard domain, in the case analyzed in the present paper, is obtained by $6$ different copies of the small billiard domain.\\
\\
\section{Quantum regime\label{section4}}
The quantum version of the billiard model is obtained by promoting to operators the momenta $\pi_a$'s in the kinetic part of the Hamiltonian $H_0=\tfrac{1}{2}G^{ab}\pi_a\pi_b$, with the metric $G_{ab}$ defined in Eq. (\ref{betametric}). As a result \cite{Kleinschmidt:2009cv}, on the UPHP, the following eigenvalue problem \cite{terras} is obtained\\
\begin{equation}\label{eigen}
-\Delta_{\rm LB}\Psi(z)=E\Psi(z)
\end{equation}
where $\Delta_{\rm LB}$ is the Laplace-Beltrami operator on the PUHP, and $\Psi$ is the wave function solving the eigenvalue problem. From a cosmological point of view, the wave function $\Psi$ can be interpreted as the wave function of the universe, in the asymptotic regime when the universe approaches the cosmological singularity, and quantum effects have to be taken into account.\\
The eigenfunctions $\Psi$ can be analyzed according to their Fourier expansion for the periodic variable $u$ \cite{terras}.\\ Maass wavefunctions $f(z)$, whose Fourier expansion $\Psi=\sum_s \Phi_s$ reads
\begin{equation}\label{maass}
\Phi_s(u,v; c_1, c_2)=c_1v^s+c_2v^{1-s}+\sum_\mu c_\mu y^{1/2}\mathcal{K}_{s-\tfrac{1}{2}}(2\pi\mid \mu\mid v)exp[2\pi i\mu u],
\end{equation}
are the only (known) eigenfunctions for the Laplace-Beltrami operator in Eq. (\ref{eigen}) invariant under the modular group. The functions $\mathcal{K}$ are the K-Bessel functions. The coefficient $c_\mu$ in Eq. (\ref{maass}) can be imposed conditions for the well-behaved-ness of the wave function with respect to the polynomial growth of $\mathcal{K}$ for $z\rightarrow\infty$. The Maass wave function (\ref{maass}) then is defined by the energy level $s$ and by the two constants $c_1$ and $c_2$.\\ 
The eigenvalue problem (\ref{eigen}) for the Fourier-expanded Maass waveforms (\ref{maass}) is then
\begin{equation}
-\Delta_{\rm LB}\Phi_s(u,v)=E_s\Psi_s(u,v),
\end{equation} 
where the eigenvalues $E_s$ defined the energy spectrum of the quantum Hamiltonian, and explicitly read
\begin{equation}
E_s\equiv s(s-1).
\end{equation}
 
Maass wavefunctions solve the problem (\ref{eigen}), and obey the (invariant) transformation law $f(\gamma z)=f(z)$ $\forall \gamma \in SL(2,Z)$.\\
Moreover, according to the features of the Fourier expansion for the $u$ variable, they can be classified as $u$-odd and $u$-even, according to their parity under the transformation $u\rightarrow-u$. In particular, $u$-odd Maass wavefunctions contain only $\sin(2\pi\mu u)$ in their Fourier expansion, while $u$-even one only $\cos(2\pi\mu u)$.\\
Finally, the behavior of Maass wavefunctions under reflections has been studied according to the parity under the reflection $R_1$ (form which the behavior under reflections in general can be obtained). Accordingly, the Maass wavefunctions can be classified according to their $R_1$-parity \cite{Kleinschmidt:2010bk}.\\ 
The quantum eigenvalue problem (\ref{eigen}) will be analyzed from two different points of view: $i$) how to impose the proper boundary conditions according to the geometrical 'shape' of the billiard table; and $ii$) how to interpret the invariance of the dynamics under the billiard maps.
Analyzing the task in two different steps will be useful in further illustrating the differences between the geometrical symmetries of the billiard tables and the features of their dynamics.\\
The following reasoning is in principle valid for both Dirichlet or Neumann boundary conditions. The choice of Dirichlet \cite{csordas1991}, \cite{Benini:2006xu} \cite{Kleinschmidt:2009hv} or Neumann \cite{Forte:2008jr} \cite{Graham:1990jd} boundary conditions for this particular problem and their comparison with respect to their quantum-mechanical interpretation has already been compared \cite{graham1991} and discussed \cite{Kleinschmidt:2010bk} in the literature. Anyhow, for the sake of compactness, only the Dirichlet problem will be mentioned; the way to treat the Neumann problem is analogous, and the comparison will be briefly discussed.    
\subsection{Empty billiards} 
Boundary conditions for the billiard table alone can be analyzed according to the 'shape' of the domains.\\
For the 'empty' unquotiented big billiard, for Dirichlet boundary conditions, the eigenfunction $\Phi$ must vanish on the three boundaries\footnote{Although the eigenvalue problem (\ref{eigen}) on the big-billiard table is well-defined, and although it is possible to treat it explicitly, in the literature other approaches are present, in which approximated or deformed domains are considered \cite{Graham:1990jd} \cite{Benini:2006xu}}, without any particular prescription on the dynamics.\\
 Automorphic forms $f(z)$ that admit a Fourier expansion
\begin{equation}
f(z)=\sum_nf_n(v)exp[i\pi \nu]
\end{equation}
can be considered as eigenfunctions of the problem (\ref{eigen}), as their \textit{odd} part would contain only the $\sin(n\pi u)$ functions, while their \textit{even} part only $\cos(n\pi u)$, for Dirichlet or Neumann boundary conditions, respectively.\\
 We can therefore take for $\Psi$ \textit{any} automorphic form $F$, solution of Eq. (\ref{eigen}), vanishing, say, on $u=0$, for the Dirichlet problem. In fact, \textit{without} requesting any identification about the remaining two sides, it is straightforward to verify, from a geometrical point of view, that the line $u=0$ is mapped into the line $u=-1$ (corresponding to the side $b$) by the transformation $z\rightarrow z-1\equiv T^{-1}(z)$, and that the line $u=0$ is mapped into the circle $u(u-1)+v^2=0$ (corresponding to the side $c$) by the transformation $z\rightarrow -z/(z-1)\equiv STS(z)$, such that, given $f(u=0,v)=0$, $\forall v$, then the proper Dirichlet boundary conditions follow, i.e.
\begin{equation}\label{geo}
f(T^{-1}(u=0,v))\equiv0\equiv f(STS(u=0,v))
\end{equation}
These eigenfunctions can be further classified according to their behavior on the 'symmetry lines' of the big billiard.\\ 
\\
Boundary conditions on the small billiard table can also be imposed \textit{without} any identification of the boundaries (which is indeed not possible), but according to simple considerations about the symmetry of the problem.\\
In more detail, the appropriate wave functions for the 'empty' small billiard can be obtained by picking up from the $u$-odd wave functions of the 'empty' big billiard those that are antisymmetric (i.e. those that vanish) on the symmetry lines and that satisfy proper Dirichlet conditions.\\
\\
As an example, we can choose, as in \cite{Graham:1994dr},
\begin{equation}\label{auto}
f(u,v)=\sum_\nu C_\nu\sin(\pi \mu u) \sqrt{v} \mathcal{K}_{s-\tfrac{1}{2}}(\pi\mid\nu\mid v),
\end{equation} 
such that the Dirichlet boundary conditions for the empty big billiard hold, according to Eq. (\ref{geo}). From these functions, one can further select those that vanish on the 'symmetry lines' of the big billiard, i.e. for $\mu=2\nu$, 
\begin{equation}\label{maassbound}
\Phi_{s}(u,v)=\sum_\mu c_\mu\sin(2\mu\pi u) \sqrt{v} \mathcal{K}_{s-\tfrac{1}{2}}(2\mu\mid\pi\mid v)\equiv\Phi_s(u,v,;c_1=c_2=0):
\end{equation} 
the odd Maass wavefunctions (\ref{maass}) are immediately obtained from Eq. (\ref{maass}) by choosing $c_1=c_2=0$ (for which cusp forms are defined), this choice being imposed by the request that the wavefunctions vanish on the boundaries.\\
\\
So far, one could be tempted to conclude that the wave functions of the small billiard consist only of a part (half) of those of the big billiard, when only boundary conditions are taken into account, according to the fact that the small billiard is a desymmetrized version of the big billiard, or, going the other way round, the big billiard consists of six (suitably glued) copies of the small billiard. However, the analysis of the dynamics will impose much stricter constraints, and will be able indeed to show their complete equivalence.

\subsection{The billiard ball} 
After having analyzed the boundary conditions that can be imposed according to the geometrical 'shape' of the billiard tables, it is possible to investigate how the structures worked out in Section \ref{section3} impose further constraints.\\
\\
Taking into account the big billiard map $\mathcal{T}$, the quotiented big-billiard maps and the small-billiard map $t$, one has to impose that the dynamics be invariant under this maps in any point, and not only on the boundaries. Furthermore, the maps $\mathcal{M}$ on the UPHP impose further constraints on the realization of boundary conditions via the proper identifications of the sides. One is therefore forced to choose Maass wavefunctions (according to their properties of being invariant under the elements of the group $SL(2,Z)$, to their $u$-parity and to their behavior under reflections) also for the big billiard both in the unquotiented case and in the quotiented ones.\\
So far, distinctions have to be taken into account.\\
The wavefunctions of the unquotiented big billiard have to satisfy suitable boundary conditions (without identification of the boundaries), and must have the appropriate behavior under the big-billiard map $\mathcal{T}$ Eq. (\ref{bbz}).\\
The wavefunctions for the BKL epoch map must obey proper boundary conditions (with the identification of the boundaries $b\rightarrow a$ and $c\rightarrow c$), and must be invariant under the transformation $T^{-1}$. In fact, the BKL epoch map contains only an even number of reflections. Of course, this paradigm is valid only for a given era, i.e. for a finite number of epochs, and does not correspond to the description of the complete dynamics.\\
The wavefunctions for the complete BKL map (i.e. when also the era-transition map is considered) and for the CB-LKSKS must obey suitable boundary conditions (with identification of the boundaries $b\rightarrow a$ and $c\rightarrow a$), and must have the proper behavior under reflections.\\
The wavefunctions for the small billiard can be obtained in principle, as explained in the discussion of boundary conditions alone, from those of the unquotiented case by 'picking up' those that vanish along the sides $B$ and $R$ (that can be interpreted as 'symmetry lines' for the big billiard). But, as already mentioned, the invariance under the maps forces one to consider only Maass wavefunctions for the unquotiented big billiard as well. As a result, because of the features of the Maass wavefunctions, the wavefunctions of the small billiard \textit{coincide exactly} with those of the full unquotiented billiard. As the big-billiard map (\ref{bbz}) consists of a suitable composition of the small-billiard map (\ref{smalltrasf}), the behavior under reflections will be the same.\\
As a result, the wavefunctions of all the kinds of $4=3+1$ cosmological billiards are the same, i.e. Maass waveforms.\\    
\\
Maass wavefunctions can be classified as odd or even according to their behavior under $u\rightarrow-u$, i.e. according to their $u$-parity. The choice (\ref{maass}) corresponds then to $u$-odd Maass wavefunctions for Dirichlet boundary conditions. Furthermore, Maass wavefunctions can be classified also according to their behavior under reflections, i.e. $R-1$-odd or $R_1$-even under the fundamental reflection $R_1: z\rightarrow-\bar{z}$. The behavior under other reflections can be straightforward analyzed; as an example, the reflection $R_2: z\rightarrow z''\equiv-\bar{z}+1$ can be analyzed as for $z'=T^{-1}z=z-1$, such that $z''=R_1(z')$. $R_1$-odd Maass wavefunctions would therefore acquire a minus sign under the action of the billiard maps (which contain all an odd number of reflections).
\subsection{Comparison of geometrical properties and dynamical features}
As already pointed out, the invariance of the dynamics of the system under the action of the billiard map follows from the existence of a reduced symplectic form, whose integral over an appropriate region (of the $(u^-u^+)$ plane) accounts for the corresponding probability for a given sequence of epochs to happen. From the quantum point of view, the probability is obtained by integrating over a suitable domain the squared modulus of the wave functions, for which a difference in sign does not produce any effect. More generally, physical observables are constructed from a suitable scalar product of wavefunctions. Accordingly, although the wavefunctions can have an even or odd behavior under reflections, this behavior is not relevant in the definition of probabilities \cite{Kleinschmidt:2010bk}. The same applies for $u$-parity.\\
Similarly, as already pointed out in \cite{Kleinschmidt:2010bk} for the discussion of the choice of boundary conditions, Dirichlet and Neumann boundary conditions produce the same expression for the probability, as complex phases would cancel out.\\
Nevertheless, it is worth remarking that a crucial difference between Dirichlet and Neumann boundary conditions arises. In the case of Neumann boundary conditions, i.e. for even Maass wavefunctions, also a \textit{continuum} spectrum appears.\\ 
Although the properties of the problem (\ref{eigen}) are already known, the comparison of the results for boundary conditions in the two different cases, i.e. the 'empty' billiard and the 'dynamical' billiard, is powerful in outlining the role of statistical billiard maps also in the quantum regime. In fact, as already pointed out, the billiard dynamics consists of a succession of free-flight Kasner trajectories joined at the reflections on the boundaries. In this evolution, the statistical maps encode all the information in the Poincar\'e return map on (a preferred) side of the billiard. As a result, the \textit{continuous} dynamics of the billiard is fully described by \textit{discrete} maps.\\
At the quantum level, it is then interesting to remark that the 'empty' boundary conditions account only for the geometrical shape of the billiard table, while the 'dynamical' boundary conditions (worked out of the statistical maps as the proper boundary identifications under the symmetry-quotienting mechanisms of the Kasner exponents) take into account the complete dynamics of the billiard (i.e. also the free-flight Kasner evolution) by imposing further constraints on the boundary conditions.\\
\\
Collecting all the ingredients together, it is possible to conclude that $i$) no complete symmetry-quotienting of the big billiard is able to eliminate reflections from the billiard dynamics. Anyhow, it is possible to identify in the BKL map two different symmetry-quotienting mechanisms, i.e. one that does not require the presence of reflections, and one that does. This difference implies a (formal) difference in the imposition of 'dynamical' boundary conditions, the difference being rather formal because of the invariance properties of Maass waveforms; $ii$) the choice of Dirichlet or Neumann boundary conditions does not affect the definition of probabilities, but the independence of probabilities from the behavior of Maass wavefunctions for $R_1$-parity and $u$-parity is somehow 'accidental', i.e. it is due to the explicit Fourier decomposition of the wavefunctions, and does not allow one in any manner to construct billiard maps which do not contain reflections (or which contain an even number of 
reflections); and $iii$) considering symmetry (sub)groups generated by transformations which do not contain reflections can be anyhow useful to understand the the symmetries and the possible identifications of the sides for cosmological-billiard tables.\\
\\
%%%%%%%%%%%%%%%%%%%%%%%%%%%%%%%%%%%%%%%%%%%%%%%

Once the BKL map has been defined for the $z$ variable on the UPHP, it is possible to determine subdomains on the big billiard table, that correspond to the first epoch of each BKL era, and to analyze the corresponding conserved quantities.\\
The definition of the subdomains of the billiard table for BKL epochs and eras follows from transferring the information gained in the ($u^-,u^+$) plane to the UPHP.\\
In the meanwhile, the dynamics of each $xy$ phenomenon covers the complete billiard table. The possibility to uniquely transfer the information to the UPHP is given by the BKL maps, as they allow one to choose a preferred ($ba$) kind of phenomena an to interpret them uniquely as fully representative of the unquotiented dynamics. Because of this, the definition of the subdomains of the billiard table by segmenting the $u$ axis according to the BKL map is fully consistent. Nevertheless, these subdomains are overlapping and do not 'tile' the big billiard table uniformly.\\ 
\subsection{BKL probabilities on the UPHP}
In the reduced phase space $(u^-,u^+)$, according to the paradigm implemented in \cite{Damour:2010sz}, it is possible to define the probability $P_{n_{xy}}$ for a $n_{xy}$-epoch era of the $xy$ type to take place as the integral of the reduced form $\omega$ in Eq. (\ref{omega}) over the corresponding box in $F_{xy}$.\\ According to the invariance properties of the reduced form $\omega$, and to the implementation of the BKL map, this probability equals the probability $P_n$ for a $n$-epoch BKL era to take place (expressed, for technical reasons, in $F_{ba}$). The integration extrema can be defined \textit{independently} for $u^+$ and $u^-$ (as the integration domains are boxes), by selecting the range of $u^-$ and $u^+$ corresponding to $n$.\\
Furthermore, given a point $u=(u^-,u^+)\in B_{ba}$, one can interpret the corresponding trajectory as the $k$-th epoch of a $(n\ge k)$-epoch BKL era, whose probability to take place equals the BKL probability $P_n$ (by the invariance of the form $\omega$ under the BKL map). Moreover, the probability $P_{n_1,n_2}$ for a $n_1$-epoch BKL era to take place and to be followed by a $n_2$-epoch BKL era composes by means of the definition of further sub-boxes in $F_{ba}$.\\ 
In particular, one can define the ranges of $u^-$ and $u^+$ as a function of the features $N$ of the epochs, i.e. 
\begin{equation}\label{ranges}
u^+_m(N)<u^+(N)<u^+_M(N), \ \ \ \ u^-_m(N)<u^-(N)<u^-_M(N),
\end{equation}
where the subscript $m$ generically denotes the minimum value, while the subscript $M$ generically denotes the maximum value. As a result, the (normalized) probability $P_N$ is given by
\begin{equation}\label{probability}
P_N\equiv \frac{1}{\ln2}\int^{u^-_M(N)}_{u^-_m(N)}du^-\int^{u^+_M(N)}_{u^+_m(N)}du^+\frac{1}{(u^+-u^-)^2}.
\end{equation}
The details of the ranges (\ref{ranges}) for the three situations described in the above and their BKL probabilities are listed in the Left Panel of Table \ref{table7}. In the Right Panel of Table \ref{table7}, for a given BKL era, the values of the radius $r$ and of the center $u_0$ of the geodesics (\ref{geod}) parameterizing the corresponding BKL epoch are listed (for future purposes).\\
It is now useful to remark that the probability $P_n$ for an $n$-epoch BKL era to take place corresponds to the area of the corresponding starting box on the ($u^-,u^+$) plane, according to the measure $1/(u^+-u^-)^2$.\\
%%%%%%%%%%%%%%%%%%%%%%%%%%%%%%%%%%%%%%%
\begin{table}
\begin{center}
    \begin{tabular}{ | l | l | l | l || l || l | l | l | }
    \hline
    $u^+_m$ & $u^+_M$ & $u^-_m$ & $u^-_M$ & $N$ & $r_m$ & $r_M$ & $u_0$ \\ \hline
    $n-1$ & $n$ & $-2$ & $-1$  & $n$ & $n/2$ & $(n+2)/2$ & $(n-2)/2$ \\ \hline
    $n-n'$ & $n-n'+1$ & $-n'-1$ & $-n'$ & $n'\le n$ &  $n/2$ & $(n+2)/2$ & $(n-2n')/2$  \\ \hline
    
    \end{tabular}
\end{center}
\caption{\label{table7} The boundaries of the subdomains, which correspond to a given BKL epoch, specified by $N$. $N=n$ denotes the first epoch of a BKL era, while $N=n'\le n$ denotes the $n'$-th epoch of a BKL era.\newline
\textbf{Left panel}. The boundaries of the boxes in the $(u^-,u^+)$ plane.\newline 
\textbf{Right panel}. The corresponding values for the radius $r$ and for the center $u_0$ of the geodesics parametrizing the BKL epoch in the UPHP.}
\end{table} 
%%%%%%%%%%%%%%%%%%%%%%%%%%%%%%%%%%%%%% 
One needs now to identify the proper region of the UPHP ( i.e. the subdomain of the big-billiard table) which corresponds to the proper box of the $(u^-,u^+)$ plane.\\
It is interesting to stress again that one of the most useful features of the reduced-phase space method is the possibility to define suitable regions and subregion in the ($u^-,u^+$) plane, which are delimited by straight lines (boxes). According to this, the definition of the allowed subdomains for epochs and eras acquires a straightforward representation. On the contrary, on the UPHP, the definition of the corresponding subregions would appear technically more complicated. This way, the best approach is to transfer to the UPHP the information gained in the reduced phase space. By means of the BKL map on the UPHP, the definition of these subdomains will acquire a well-posed meaning. In fact, one should remember that each $B_{xy}$ region of the ($u^-,u^+$) plane covers the complete billiard table.\\
\\
Transferring the information available in the reduced phase space, as listed Table \ref{table7}, one learns that
 these region will be delimited by $-1<u<0$ and by the proper specifications for the trajectories of the billiard parametrized by geodesics $\mathcal{G}$, as in Eq. (\ref{geod}). Following the notation of Eq. (\ref{ranges}), one then finds the integration extrema by plugging the results recalled in the Right Panel of Table \ref{table7} into Eq. (\ref{geod}). In particular, the range $v^m(n)<v<v^M(n)$ has to be worked out of Eq. (\ref{geod}).\\
One then finds that the domains $D_n$ correspond to the portion of the billiard table between two geodesics (\ref{geod}) $\mathcal{G}^M$ and $\mathcal{G}^m$, defined as
\begin{subequations}\label{asymptile}
\begin{align}
&\mathcal{G}_M\equiv\mathcal{G}_M(u^-_m, u^+_M),
&\mathcal{G}_m\equiv\mathcal{G}_m(u^-_M, u^+_m).
\end{align}
\end{subequations}
and the 'vertical' walls of the billiard, $u=0$ and $u=-1$. The area $a_n$ of each subdomain $D_n$ is evaluated according to the measure $da$ on the UPHP,
i.e. 
\begin{equation}
da=\frac{dudv}{v^2},
\end{equation}
such that the area $a_n$ of the subdomain corresponding to the $n$-epoch BKL era is given by
\begin{equation}\label{areaen}
a_n=\int_{-1}^0du\int^{v_M(n)}_{v_m(n)}\frac{dv}{v^2}
\end{equation}
Differently form the case of the $(u^-, u^+)$ phase space, the geodesics that 'pave' the UPHP do not form a perfect tiling, but it is possible to recognize that, in the limit for large $N$ in Table \ref{table7}, the circles $\mathcal{G}_M$ and $\mathcal{G}_m$ in Eq. (\ref{asymptile}) can be approximated by 'straight horizontal lines' in the billiard table.\\ 
As one verifies straightforward, the measure $dudv/v^2$, i.e. the invariant measure on the UPHP, is invariant under the BKL epoch map and under the CB-LKSKS era transition map on the UPHP. As a result, the quantity $a_n$ is conserved during the evolution of the billiard. \\
One can also interpret Eq. (\ref{areaen}) as the integral of the area form $\mathcal{A}(u,v)$, i.e.
\begin{equation}\label{areaform}
\mathcal{A}(u,v)=\frac{du\wedge dv}{v^2}.
\end{equation}
which is therefore a form invariant under the BKL map and the CB-LKSKS map. Its integral over suitable domains will provide conserved quantities. The invariant form (\ref{areaform}) can also be considered to be obtained form the invariant $\omega$ form on the UPHP, when a suitable cross section of the Hamiltonian flow is taken into account, as stressed in \cite{Damour:2010sz}.\\  
\\
The conserved quantities $\tilde{a}_n(c_1,c_2;s)$
\begin{equation}\label{areaenm}
\tilde{a}_n(c_1,c_2;s)=\int_{-1}^0du\int^{v_M(n)}_{v_m(n)}\mid\Phi_s(u,v; c_1,c_2)\mid^2\frac{dv}{v^2}
\end{equation}
are invarian under the BKL epoch map and the CB-LKSKS map are evaluated on the regions of the UPHP which correspond to a given BKL epoch, and are obtained by transferring on the UPHP the information gained in the restricted phase-space. They are obtained from the $s$-th Fourier mode of the decomposition $\Phi_s$ of the wave function $\Psi$, and depend on the number $n$ of the BKL map. 
It is interesting to remark that, from a \textit{classical} point of view, one does not need to require that the coefficients $c_1$ and $c_2$ in Eq. (\ref{maass}) and in Eq. (\ref{areaenm}) vanish. In fact, from the classical point of view, one only needs invariance under the BKL epoch map and the CB-LKSKS maps. The expression (\ref{maassbound}) is obtained, in the quantum regime, by imposing boundary conditions.\\
\\
 %%%%%%%%%%%%%%%%%%%%%%%%%%%%%%%%%%% 
\section{Quantum BKL maps\label{section4a}}
One has now collected all the features of the quantum regime that allow one to express (at least formally) the 'quantum BKL probability' for epochs and eras to take place.\\
\\    
After the implementation of the quantum regime, the expression for the quantum probabilities, as the integral of the squared modulus of the Maass waveform over the suitable domain in the UPHP, follow directly. In fact, this has the direct physical interpretation of the quantum probability for the billiard ball to be in the corresponding subregion. By means of the BKL map on the UPHP, and of the division of the billiard table into subdomains $D_n$ which correspond to an $n$-epoch BKL era, it is possible to obtain the quantum analogue of the classical BKL probabilities. \\
Moreover, the invariance of the Maass wavefunctions under the BKL map will ensure the correct behavior of these probabilities during a BKL era.\\  
\\
\subsection{Quantum implementation of probabilities for BKL epochs and eras.}
As anticipated in the above, it is now possible to establish a formal definition for the quantum version of BKL probabilities for a $n$-epoch BKL era to take place, say $\textbf{P}_{n,s}$. \\
As a result, the quantum BKL probability $\textbf{P}_n$ will be given by the probability for the billiard ball to be found into the proper integration domain in the billiard table. In other words, one has to look for the probability that a geodesics $\mathcal{G}$ (parameterizing a BKL epoch) be between the proper integration domain.\\
\\
\subsection{Formal results}
Quantum BKL probabilities $\textbf{P}_N$ will then be given by the probability for the billiard ball to be found between the proper integration domain $D_n$ in the billiard table. They are expressed as the integral of the squared modulus of the Maass waveform (\ref{maassbound}) over the proper integration domain $D_n$, such that
\begin{equation}\label{quantumbkl}
{\bf P}_{n,s}=\int_{-1}^0du\int^{v_M(n)}_{v_m(n)}\frac{dv}{v^2}\mid\Phi_{s}(u,v)\mid^2.
\end{equation}
In Eq. (\ref{quantumbkl}), the wave function is $\Phi_s(u,v; c_1=c_2=0)$ in Eq. (\ref{maassbound}) rather than that of Eq. (\ref{maass}), because the proper boundary conditions have to be recovered.\\
\\
\subsection{Physical interpretation of the quantum probabilities with respect to the BKL maps}
Given the formal result (\ref{quantumbkl}), the physical interpretation of the quantum BKL probabilities can be understood.\\
\\   
At the classical level, the reduced $\omega$ form is obtained by choosing \textit{ab initio} a fixed energy shell, and then by considering the Poincar\'e return map on the sides of the billiard.\\
The BKL probabilities are defined as the integral of the reduced $\omega$ form on the pertinent regions of the restricted phase space. Transferring the corresponding information from the restricted phase space to the UPHP, via the quotienting of the roles of the Kasner exponents in the BKL map, one finds the corresponding subregions of the big billiard table by choosing the correct geodesics from their endpoints on the $u$-axis.\\
\\     
On the other hand, the dependence of the quantum BKL probability on the energy level has therefore no straightforward classical analogue. In fact, the eigenvalue problem (\ref{eigen}) admits a complete spectrum for the energy levels $E_s$, and it is not possible to fix any value for the energy, and restrict the available phase space accordingly. On the contrary, it is however possible to define a quantum BKL probability for each energy level. As a result, the quantum BKL probabilities 'scan' the complete energy spectrum of $E_s$ and receive a contribution from each energy level $s$. This contribution is exactly the contribution of the generalized invariants (for a particular choice of the constants $c_1$ and $c_2$).\\ 
\\
Moreover, the quantum BKL probabilities (\ref{quantumbkl}) inherit from the Fourier expansion of the Maass waveforms for the periodic variable $u$ also the sum on the $\mu$ index of the $u$ periods. As a result, the quantum BKL probabilities 'project' the information about the length $n$ of a BKL era along the directions $\mu$ of the $u$-periodic expansion. In other words, a sort of (generalized) Fourier expansions in the $\mu$ directions is obtained also for the values $n$, which defines the BKL length of an era.\\      
\\
Given all these considerations, the quantum BKL probabilities are characterized, for each length $n$ of the BKL eras, by 'sampling' all the energy levels of the eigenvalue problem (\ref{eigen}) via the corresponding generalized invariant, and by projecting them along the $\mu$ directions of the Fourier decomposition for the variable $u$.\\
\\
\subsection{More about the UPHP}
Furthermore, given all the features of the quantum version of BKL probabilities, it is possible to comment on the generalized area invariants (\ref{quantumbkl}).\\
In fact, as already mentioned, the generalized area forms do not owe their invariance to the choice of a fixed energy shell. The corresponding conserved quantities are defined for domains $D_n$, which have been defined according to a fixed energy shell, to which $\omega$ owes its invariance in the reduced phase space. Accordingly, one can interpret the occurrence of all the $s$ generalized forms as compensatory for all the possible classical values of the (classical) Hamiltonians, which have been integrated away to obtain $\omega$ itself.\\
On the other hand, from the classical point of view, the Maass functions (\ref{maass}) are defined as invariant under the elements of the pertinent transformation group. In principle they can be investigated also without taking into account the eigenvalue problem; or, more in general, the eigenvalue problem can be investigated regardless to the meaning it acquires in quantum mechanics. But, in the quantum regime, the spectrum of the Laplace-Beltrami operator on the UPHP acquires the meaning of the succession of energy levels for a 'particle in a box' (i.e. the quantum billiard dynamics). This way, choosing real eigenvalues for the Schroedinger equation in the quantum regime implies the interpretation of the class of generalized area invariants on the UPHP as due to the \textit{real} possible values for the (conserved) energy of the (classical) Hamiltonian, which are real.\\
\\
\subsection{Open issues}
It is in order to remark that the definition of quantum BKL epochs and eras requires further investigation.\\
On the one hand, the renormalizability of the integrals defining the quantum BKL properties on the UPHP, where suitable subregions of the billiard table have to be taken into account, seems to require numerical simulation.\\
On the other hand, the role of the quantum BKL properties in the definition of the quantum nature of the cosmological singularity has not been clarified yet. In fact, in the pioneering paper by Misner \cite{misn}, the billiard table is examined on the unit circle rather than on the UPHP, and the quantum numbers that are hypothesized to describe the quantum nature of the cosmological singularity are obtained by considering a squared potential well, rather than the exact triangular domain. Moreover, it is possible to demonstrate that the Misner variables correspond only asymptotically (towards the singularity) to the description on the UPHP.\\
On the other hand, from the classical point of view, the BKL length of epochs and eras is crucial in the definition of the chaotic nature of the cosmological singularity, when an isotropic and inhomogeneous universe is considered. In this respect, no previous interpretation of its quantum nature has been proposed.\\ 
\\
Furthermore, the role of the energy shells defined by the Fourier modes $\mu$ of the decomposition of the wave function $\Phi_s$ has no classical analogue, and can be interpreted as the possibility to isolate only selected energy shells. In this respect, it is necessary to stress that the definition of BKL probabilities from the classical point of view is connected with the billiard description of the cosmological singularity on by fixing ab initio a particular energy shell in the Hamiltonian flow \cite{Damour:2010sz}; this suggests that the quantum definition of these energy shells is not connected with the quantum description of billiards on the hyperbolic plane in general, but descends only from the invariance of the  the proper integrals of the Maass waveform on the UPHP under the BKL map. Also in this direction, further investigation is needed to clarify the role of (a suitable selection of) energy shells in the description of the quantum nature of the universe.\\
Nevertheless, the presence of these new structures in the wave function on the UPHP has to be regarded to as the discovery of new structures in the quantum nature of the cosmological singularity. Indeed, the renormalizability properties of BKL probabilities in the classical framework depends crucially on the choice of a fixed energy shell. In the quantum case, these properties has to be studied yet. Nonetheless, the case of a non-renormalizable probability set due to a non-standard measure is not to be considered as a flaw in the construction; in fact, also in \cite{Damour:2010sz}, a non renormalizable measure on the unit circle has been defined.\\
\\
It is worth remarking that the results here obtained about the quantum BKL maps and the quantum BKL numbers are entirely based on the symmetries of the wavefunction which solves the eigenvalue equation for the Laplacian operator on the UPHP, according to the symmetries of the model. The definition of such maps is then tailored around the features of cosmological billiards in $4=3+1$ dimensions, and is effective in matching the properties of both the big billiard and the small billiard, which account not only for a very precise physical interpretation of these models as far as the early stages of the evolution of the universe are concerned, but also provide a mathematically well-defined $4$-dimensional limit for all those higher-dimensional models, for which a billiard map of the solution to the field equations is obtained, and for which new more complicated structures are discovered in the higher-dimensional versions
\cite{Damour:2002fz}, \cite{Damour:2002cu}, \cite{Damour:2001sa}, \cite{Damour:2000hv}, \cite{Fre':2005si}, \cite{Fre:2005bs}, \cite{{damnew}}.\\ 
Some of the processes which characterize the BLK maps and the CB-LKSKS maps at the classical regime have strong physical implications. The mathematical definition of these features for the quantum version of these maps, as well as for their semiclassical limit should therefore be centered on the physical information which is encoded. In fact, the definition of quantum BKL maps and of quantum CB-LKSKS maps has to be based on the symmetries of the billiard tables, those of the classical dynamics, and have to automatically contain the geometrical features and the mathematical structures which characterize these stage of the evolution of the universe within the BKL paradigm. The general features of quantum maps and of wavefunctions, in the quantum scheme, in the semiclassical limit and at the classical level as far as the presence of chaos is concernd are discussed in \cite{berry}, where the semiclassical regime is recognized at scales corresponding to the de Broglie wavelength. A definition of quantum maps from the definition of the action of suitable operators was formulated in \cite{cairoli}.\\
The implementation of precise statistical maps to the quantum regime requires technical skills and interpretative efforts, according to the phenomena schematized by the particular maps. Fos some kinds of maps which have some common features with the Gauss map, a large amount of investigation has already been produced. As an example, it is worth recalling that a quantum version of the Baker's transformation has been formulated in \cite{balvor1989} and in \cite{rubin}. The examination of the role played by the Planck's constant within the establishment of the quantum version of statistical maps, such as in \cite{leray}, is effective in pointing out the relevance of the physical interpretation of these procedures in general, and of the physical content of the BKL paradigm for the case of cosmological billiards, also in the case a different number of space-time dimensions is considered. The relation between classical trajectories and quantum ones within the framework of the quantum Baker's map has been defined in \cite{inoue}. The semiclassical limit of these maps has been addressed in \cite{kaplan} , and in \cite{keating} the semiclassical limit of a sequence of eigenstates has been approached. The semiclassical limit of the iteration of a quantum statistical map has been defined in \cite{saraceno}. A definition of the quantum Bernoulli map has been proposed in \cite{ordonez} as derived form the quantum Baker's map; the relation between the two maps has been proved in \cite{soklakov} as far as the Planck's constant is involved. The role of the Poincar\'{e} recurrence theorem in the semiclassical expression of a statistical map (in the case of the Birkhoff's map) has been discussed in \cite{vallejos}.\\ 
After the identification of some of the problems which affect the definition of a quantum statistical maps, as some of the properties exhibited in the classical regime can be lost, or their quantum implementation can bring interpretative mismatches, it is interesting to remark that the definition of quantum BKL maps and of quantum BKL configurations, and their associated conserved quantities, are effective in maintaining the role of the statistical maps in encoding the information gained by the discrete Poincar\'{e} return map on the boundaries of the billiard table, in defining the symmetries under which quantum states must be invariant, for which the iteration of the quantum version of the maps has a well-defined quantum version and semiclassical limit, for which a symbolic description of the dynamics by means of the succession of bounces on the sides of the billiard table is encoded in the sequence of epochs and eras.\\
Furthermore, the analysis of these maps allows one to establish the precise requirements which must be fulfilled by a definition of the asymptotic properties of the wavefunctions, when a quantum version of gravity has to be envisaged, and when the definition of such models implies any modification of the geometry of space-time. Indeed, the symmetries of the wavefunctions, which encode the chaotic properties of the dynamics and  not only the symmetries of the billiard tables, have to be encoded also in any phenomenon, for which chaos would in principle be removed.\\
%%%%%%%%%%%%%%%%%%%%%%%%%%%%%%%%%%%%%%%%%%%%%%%%%%%%
\section{Discussion of the results\label{comparison}}
In this Section, the main steps in the definitions of statistical maps for cosmological billiards and for the shape and the symmetries of their domains and their dynamics are recalled and compared.\\
\\
The original one variable BKL map for the variable $u$ that parameterizes the solution of the Einstein field equations in the asymptotic limit towards the cosmological singularity, according to billiard interpretation of the dynamics, was given in (5.4) of \cite{BLK1971}.\\
\\
The two-variable CB-LKSKS map was given in the first part of Eq. (4) of \cite{Chernoff:1983zz}, while in \cite{sinai83} and in Section 4 of \cite{sinai85}, for the two-variable map of the unit square to the unit square.\\
\\
In \cite{Damour:2010sz}, the reduced phase space technique has been developped is Section IV and in the Appendix, and used for the description of the features of cosmological billiards in $4=3+1$ spacetime dimensions.\\
The restricted phase space for the variables $u^+$ and $u^-$ is illustrated in Figure 6 and the detail are reported in Table II; the corresponding unquotiented map for the full dynamics of the big billiard is given in Eq.'s (5.2), (5.3) and (5.4).\\
The BKL epoch map is given in Table IV, while the corresponding subregions of the restricted phase space are sketched in Figure 6; the BKL era map is given in Table V, and the pertinent subregions of the restricted phase space are depicted in Figure 7.\\
The two-variable Kasner-quotiented era map, which corresponds to the two-variable CB-LKSKS map for the variables $u^+$ and $u^-$, is given in Eq. (6.2).\\
The expressions for the Kasner transformations for the restricted phase space and their properties are listed in Table I and Table VI.\\
The restricted phase space for the small billiard is established in Table VIII and represented in Figure 8.\\
The unquotiented small billiard map for the variables $u^+$ and $u^-$ is defined in Eq.s (9.3), (9.4) and (9.5).\\
The possible symmetry-quotienting mechanisms for the small billiard are discussed in Subsection IX B.\\
\\
In \cite{Fleig:2011mu}, the volume formula for the domain of the small billiard in any spacetime dimensions is found, Eq. (32).\\
The volume of the big billiard domain is shown to be determined by the index of the subgroup which defines the big billiard with respect to the small billiard in Section I (ibidem).\\ 
\\
In the present work, this investigation has been developped in the complex variable $z=u+iv$ for the UPHP.\\
The congruence subgroup which defines the big billiard domain with respect to the small billiard domain is found as $\Gamma(2)$, of index $6$, from the analysis of Eq. (\ref{bbz}) in Section \ref{congruence}.\\
The unquotiented big billiard map has been given in Eq.'s (\ref{bbz}).\\
For the symmetry-quotiented maps, the variable $z_{\rm BKL}$ has been defined in Eq. (\ref{BKLzetaepoch}).\\
The Kasner-quotiented BKL epoch map has been obtained in Eq. (\ref{bklzepoch}), for the complete complex variable $z_{\rm BKL}$, and in Eq. (\ref{bkluvepoch}) for the two directions $u$ and $v$.\\
The Kasner quotiented era map has been defined in Eq. (\ref{BKLz}) for the variable $z_{\rm BKL}$ and in Eq.s (\ref{BKLuv}) for the two variables $u$ and $v$.\\
The complex CB-LKSKS map has been established in Eq. (\ref{cblksksz}).\\
The complex Kasner transformations for the UPHP have been listed in Table \ref{table1}.\\
The unquotiented small billiard map has been defined in Eq. (\ref{smalltrasf}) for the UPHP.\\
A new partition of the reduced phase space for the small billiard has been motivated in Subsection \ref{smallbkl}, presented in Table \ref{table8} and Table \ref{table9} and illustrated in Figure \ref{illustraz3}.\\
The symmetry-quotiented BKL map for the small billiard has been defined in Eq. (\ref{tsb}).\\
To complement the reduced phase space method developped in \cite{Damour:2010sz}, the tiling of the UPHP has been presented in Table \ref{table7}.\\
The implications of these methods on the quantum regime have been proposed in Section 6 and Section 7.\\
\\
This investigation has proven to be effective in clarifying some issues of the model, which had not found yet a full characterization in the previous analysis.\\ 
In addition, the use of the complex variables allows one to construct links between the transformations which act on the coordinates, its group theoretical structure, and the physical meaning of the trajectories which map the solution of the Einstein field equations.\\
\\
The BKL epoch map for the complex variable $z=u+iv$ of the UPHP, given in Eq. (\ref{bklzepoch}) for the complex variable $z$ and in Eq. (\ref{bkluvepoch}) for the two variables $u$ and $v$ separately, generalizes the two-variable BKL epoch map for the oriented endpoints found in \cite{Damour:2010sz}. Accordingly, the BKL era-transition map, given in Eq. (\ref{BKLz}) for the complex variable $z$ and in Eq. (\ref{BKLuv}) for the variables $u$ and $v$ separately, and the CB-LKSKS maps, given in Eq. (\ref{cblksksz}), respectively, express the maps found in \cite{Damour:2010sz} for the oriented endpoints, as classified in the above, which, on their turn, are interpreted as expressing the geometrical connection between the full (unquotiented) chaotic dynamics of the early universe and the original BKL description.\\
The definition of the Kasner transformations for the UPHP, given in Table \ref{table1}, for the complex variable $z=u+iv$, define the exact transformations, which reduce to the Kasner transformations for the oriented endpoints previously found in Table I and Table VI of \cite{Damour:2010sz}.\\
The symmetries of the big billiard are connected with the symmetries of the solution of the Einstein field equations, which are based on the unordered triple of the Kasner parameters. The geometrical description of the billiard tables fixes an order for the three projective directions, such that the symmetry properties of the Einstein equations are schematized by the bounces on the billiard sides. The identification of the sides which is obtained in each of the maps, i.e. in the epoch maps and in the era map, are a consequence of the symmetries of the billiard table. These identifications are summarized in Table \ref{table5}. On the one hand, there exist several transformations, i.e. several composition of the generators of the symmetry groups which characterize the solution of the Einstein equations, which can define the same transformations of the oriented endpoints, in the reduced phase space description of the dynamics. Nevertheless, the analysis on the UPHP is effective in determining the transformations, which describe the physical trajectories, i.e. the trajectories which can be interpreted as the continuous dynamics of a frictionless billiard system on a surface of constant negative curvature.\\
\\
The symmetries of the billiard table account for the symmetries of the unordered triple of the Kasner directions, which schematize the evolution of the early universe, after the solution of the Hamiltonian constraint. On the contrary, the characterization of this phenomenon by means of the correspond Poincar\'{e} return map on the appropriate surface of section does not fully take into account the mechanism according to which the evolution of the Kasner parameters (with respect to the logarithmic time) cross. This mechanisms has been here elucidated in Table \ref{table8} and exposed in Table \ref{table9}. For these properties, the definition of a statistical map for the small billiard, i.e. in the inhomogeneous case, in the UPHP, as in Eq. (\ref{tsb}), needs the introduction of an extra reflection, as in Eq. (\ref{tsb2}) with respect to Eq. (\ref{tsb1}) for the pertinent regions of the restricted phase space for which a particular pattern of behavior of the scale factors is outlined.\\
Differently form the previous analysis, this mechanism accounts for the exact point (defined by the logarithmic time by which the evolution of the scale factors is evaluated) at which the right hand side of the Einstein field equations in the asymptotically case of the Bianchi IX universe is not negligible any more, and a reparametrization of the solutions to the field equations is needed for the approximation to be valid. As a result, the definition of the small billiard map is the exact tool by which the exact behavior of the Kasner scale factors is described. It is also crucial to remark that this map solves two delicate problems.\\
The definition of a complete statistical map for the small billiard in the reduce phase space is not possible, as the accessible regions of the phase space for this model are delimited by curvilinear domain, for which the chaotic and mixing properties of the statistical maps do not allow any (at least analytically expressed) map. As a consequence, any attempt to describe the description of the small billiards by means of an analytic relation between the number of bounces on the gravitational wall (as accounted for the definition of the Poincar\'{e} return map on the corresponding surface of section) and the value of the $u^+$ variable would be affected by the need to introduce some $+1$ or $-1$ factors at (analytically) impredictable steps of the epoch map in the vicinity of the era transition (i.e. as the value for which the $R$ symmetry walls is approached). The use of the complex variable $z$ on the UPHP completely solves this issue by the introduction of the extra reflection in Eq. (\ref{tsb2}) in the corresponding subregions of the restricted phase space, as analyzed in Table \ref{table8} and illustrated in Figure \ref{illustraz3}.
From a more general point of view, this analysis constitutes also a different effort in establishing the role of the Poincar\'{e} return map for the exact behavior of the Kasner exponents. In fact, the partition of the restricted phase space into the regions, according to which a different number of reflections is needed for the definition of a statistical map, clearly defines the limitation implied by a statistical description. In other words, this construction outlines that it is possible to define the points (with respect to the logarithmic time) where the scale factors vanish, but is is not possible to exactly predict the point (with respect to the same logarithmic time) where the non-oscillating scale factor changes its slope.\\
\\
The reduced phase space description of the cosmological billiards has been adopted in \cite{Damour:2010sz} as, in this space, the different regions, for which the exact length of era (with respect to their content of epochs, which, on their turn, implies their length with respect the logarithmic time which describes the evolution toward the cosmological singularity) are delimited by straight lines. The analysis of the same regions on the UPHP, coordinatized by complex variable, as in Table \ref{table7}, shows that the corresponding regions are indeed overlapping, and constitute an irregular tiling for the UPHP. The asymptotic (i.e. the limit for which the BKL eras contain a large number of epochs) is indeed accounted for a limiting pattern, in which the domains are overlapping, but the tiling is defined by (in the limit) straight lines.\\
\\
The quantum version of the model brings itself several fascinating interpretative possibilities. Among all the intriguing direction which could be suggested by this analysis, both  from the viewpoint of the physical meaning of the wavefuntion of the universe, as well as for the definition of the mathematical properties of the Laplace-Beltrami operator for the UPHP, two main aspects of the model have been favored, which descend form the classical analysis here developed.\\
The expression of the mathematical features of the solution to the eigenvalue equation for the Laplace-Beltrami operator in the UPHP has been connected to the functional dependence of the wavefuntion on the variables, which account for the BKL lengths, according to the tiling of the UPHP, as in Table \ref{table7}. This formal result is crucial in defining the semiclassical limit of the model, interpreted as the lowest order of the WKB expansion, for which the quantum solution are evaluated on the classical trajectories. Within this analysis, it is possible to compare the quantum wavefuntion and the most probable (with respect to the BKL statistics) phase space configurations.\\
Consequently, the wavefunctions acquire invariance properties under the transformations which define the aforementioned (in the BKL statistical interpretation) phase space variables which account for the number of epochs which define the BKL succession of eras. The classical BKL maps (\ref{bklzepoch}), (\ref{BKLz}), (\ref{cblksksz}) both for the big billiard and for the small billiard, are effective in selecting the symmetries of the wavefunctions and the boundary conditions. Furthermore, the statistical map (\ref{tsb}) for the small billiard is irreplaceable in reconciling the features of the dynamics accounted by the wavefuntion of the small billiards, for which the symmetries of the wavefunctions of the big billiard are obtained by piecewise 'gluing' the wavefunctions of the small billiard, thus obtaining the result of the pertinent congruence subgroup. This analysis is relevant in picking up the physical features of the billiard motion with respect to all the previous analysis, where the group theoretical properties had already been strictly elucidated, but not connected with the continuous billiard dynamics, which is described by the discrete BKL statistical description, and by the 'impredictable' (with respect to the discrete maps) behavior of the non-oscillating scale factor, which is described, on its turn, by the presence of an extra reflection in the small billiard map (\ref{tsb}).\\
\section{Concluding remarks\label{section5}}
In this paper, the structure underlying cosmological billiards in $4=1+3$ dimensions has been investigated. In particular, the unquotiented big billiard, the $S_3$-symmetry quotiented billiard, and the small billiard have been taken into account. The BKL map and the CB-LKSK map consist in quotienting out of the dynamics the ($6$ possible) permutations of the Kasner exponents, while the small billiard is obtained in the inhomogeneous case, and consists namely of $(1/6)$ of the big billiard.\\
\\
More in detail, the study of the dynamics on the UPHP allows one to recover the exact content of reflections in the different maps. This information is important in  testing and in defining the equivalence between the different maps, and to establish the conditions under which this equivalence holds, and to find out the reason of apparent mismatches in the comparison of the big billiard dynamics and the small billiard dynamics. In fact, the analysis of the BKL quotiented dynamics of the big billiard reveals the mechanisms according to which the BKL maps are obtained as far as their content of reflections is concerned. On the other hand, the BKL map for the small billiard consists is obtained for two different cases, which contain a different number of reflections.
The features of the BKL maps here outlined are useful in defining and in understanding the differences between the small billiard and the big billiard.
The big billiard accounts for pure gravity (i.e. gravitational walls only are taken into account), while the small billiard, which corresponds to the more generic inhomogeneous case, is delimited by two symmetry walls and  (half) gravitational wall.% The small billiard is the fundamental domain of the group PG, whose generators are the building blocks of the transformation that define the big billiard, as a suitable congruence subgroup. 
 Within this framework, the big billiard is relevant as it has one entire gravitational wall, on which the Poincar$\acute{e}$ return map is implemented. On the contrary, the small billiard contains only half gravitational wall, such that the behavior of the trajectories which are not explicitly considered for this wall are obtained as a suitable Weyl reflection.
The discrepancy between the number of reflections in the different maps is a very interesting investigation subject, which can shed light on the features of the dynamics of the billiard rather than the features of the composition of its generators.
This way, it is necessary to consider the big billiard as it is the smallest closed domain, which can be constructed by suitably 'gluing together' suitable rotation of the fundamental domain, which contains a complete gravitational wall, on which statistics are performed.\\
%Within this viewpoint, the BKL maps, which may involve one variable or two variables, are relevant in reducing the complete $4$-variable CB map \cite{Chernoff:1983zz} after the definition of a Poincar\'e surface of section, thus effectively reducing the continuous dynamics of the billiard motion to a discrete map.\\
\\
The implementation of the quantum regime has been carried out on the basis of both boundary conditions and properties of the dynamics. As a result, a difference in the imposition of the boundary conditions appears, but this difference is eliminated as soon as the dynamics is taken into account. More in detail, although a more general wave function can be chosen in the case of the unquotiented big billiard as far as boundary conditions are concerned, the Maass wavefunctions remains the only possibility when the dynamics has to be taken into account.\\
An investigation of conserved quantities under the BKL map and the CB-LKSKS map in the UPHP at the classical level allows one to introduce a quantum analogue for the BKL probabilities. The quantum version of the BKL probabilities for the length $n$ of an era has been established as the integral of the squared modulus of the eigenfunction of quantum the Hamiltonian over a suitable subdomain of the big-billiard table, and can be compared with the occurrence, at a classical level, of generalized area forms on the UPHP. As a result, the BKL probabilities experience the energy levels $s$ of the spectrum of the quantum Hamiltonian and project each contribution along the directions $\mu$ of the Fourier series of the Maass wavefunctions for the periodic variable $u$. 
\\
The work has been organized as follows.\\
In Section \ref{section1}, the main issues about cosmological billiards have been introduced.\\
In Section \ref{section2}, $4=3+1$ cosmological billiards have been obtained, and the problem of suitably quotienting the dynamics has been focused on.\\
In Section \ref{section2a}, the main features of some groups have been recalled, as they prove to be a very powerful tool in describing the dynamics of the complete big billiard, as well as those of its symmetry-quotiented versions.\\
In Section \ref{section3}, the properties of the BKL epoch map, the BKL era-transition map and the CB-LKSKS map are compared. Although for these maps the properties of the dynamics are encoded in a description that destroys the typical behavior of the billiard, consisting of bounces (reflections) against the sides of the billiard table, a difference is found between the BKL epoch map and the CB-LKSKS map, interpreted as a composition of BKL era-transition maps. This is due to the different approaches in quotienting the roles of the Kasner exponents.\\
In Section \ref{congruence}, the structure of the big billiard with respect to that of the small billiard has been established.\\
In Section \ref{section4}, particular attention has been devoted to the definition of invariant quantities on the UPHP, and the quantum implementation of the model has been addressed. More precisely, prescriptions about the the quantization are obtained considering both boundary conditions and maps.\\
In Section \ref{section4a}, a particular case of observables has been considered: the quantum versions of BKL probabilities for the length $n$ of a $BKL$ era. The quantum BKL probabilities for each $n$ have been shown to depend also on the particular energy level $s$ of the spectrum of the quantum Hamiltonian, and to project this information along the directions $\mu$ of the Fourier expansion of the Maass waveforms for the periodic variable $u$.\\
In Section \ref{comparison}, a the comparison of the achievements of this work with respect to previous opens issues has been established.\\
Section \ref{section5} contains the consluding remarks.
%%%%%%%%%%%%%%%%%%%%%%%%%%%%%%%%%%%%%%%%%%%%%%%%%%%%%%%%%%%%%%%%%%%%%%
\section*{Acknowledgments}
OML is grateful to the Albert Einstein Institute- Max Planck Institute for Gravitational Physics for warmest hospitality during the early stages of this work and during the revision of the manuscript. OML would like to thank Prof. T. Damour for the discussion of the dynamical subregions of the epoch hopscotch court, and Prof. H. Nicolai for helpful clarifications and explanations about group theory, and A. Kleinschmidt for critically revising the manuscript.\\
%%%%%%%%%%%%%%%%%%%%%%%%%%%%%%%%%%%%%%%%%%%%%%%%%%%

%%%%%%%%%%%%%%%%%%%%%%%%%%%%%%%%%%%%%%%%%%%%%%%%%
\end{document}